\documentstyle[12pt, epsfig]{article}     
     
\begin{document}     
     
\vspace*{-2cm} \renewcommand{\thefootnote}{\fnsymbol{footnote}}     
\begin{flushright}     
hep-ph/9910338\\     
DTP/99/94 \\     
PSI PR-99-24\\ 
October 1999\\     
\end{flushright}     
\vskip 65pt     
     
\begin{center}     
{\Large {\bf Resummation of double logarithms in electroweak high energy     
processes}}\\[0pt]     
\vspace{1.2cm} {\bf V.S. Fadin, ${}^1$\footnote{{\bf fadin@inp.nsk.su}} L.N.     
Lipatov, ${}^2$\footnote{{\bf lipatov@thd.pnpi.spb.ru}} A.D. Martin,     
${}^{3}$%
\footnote{{\bf A.D.Martin@durham.ac.uk}} and M. Melles ${}^4$\footnote{{\bf     
Michael.Melles@psi.ch}} }\\     
\vspace{10pt}   
  
{\sf 1) Budker Institute of Nuclear Physics and Novosibirsk State University, 630090   
Novosibirsk, Russia \\ }  
\medskip  
  
{\sf 2) St. Petersburg Nuclear Physics Institute, 188350 and St. Petersburg     
State \\  
University, St. Petersburg, Russia \\ }  
\medskip  
  
{\sf 3) Department of Physics, University of Durham, Durham, DH1 3LE, UK     
\\ }  
\medskip  
     
{\sf 4) Paul Scherrer Institute (PSI), CH-5232 Villigen, Switzerland. }  
\end{center}

\vspace{60pt}     
\begin{abstract}     
At future linear $e^+e^-$ collider experiments in the TeV range, Sudakov     
double logarithms originating from massive boson exchange can lead to     
significant corrections to the cross sections of the observable processes.     
These effects are important for the high precision objectives of the Next     
Linear Collider. We use the infrared evolution equation, based on a gauge     
invariant dispersive method, to obtain double logarithmic asymptotics of     
scattering amplitudes and discuss how it can be applied, in the case of     
broken gauge symmetry, to the Standard Model of electroweak processes. We     
discuss the double logarithmic effects to both non-radiative processes and     
to processes accompanied by soft gauge boson emission. In all cases the     
Sudakov double logarithms are found to exponentiate. We also discuss double     
logarithmic effects of a non-Sudakov type which appear in Regge-like     
processes.     
\end{abstract}

\vskip12pt     
     
\setcounter{footnote}{0} \renewcommand{\thefootnote}{\arabic{footnote}}     
     
\vfill     
\clearpage     
\setcounter{page}{1} \pagestyle{plain}     
     
\section{Introduction}     
     
The Next Linear Collider (NLC) will explore $e^{+}e^{-}$ processes in the     
TeV energy regime, and probe the Standard Model of elementary particles to     
great accuracy. Electroweak processes which may reveal New Physics, such as     
supersymmetry, are especially interesting. Therefore the accurate     
calculation of the scattering amplitudes of such high energy processes in     
the Standard Model is very important. The main corrections to the Born     
amplitudes at high energies, $\sqrt{s}$, are double logarithmic (DL)     
contributions of the form $(g^{2}\log ^{2}(s/m^{2}))^{n}$ which arise from     
soft vector boson exchanges. Here we derive the DL asymptotics of scattering     
amplitudes, for a sequence of gauge theories leading up to the Standard     
electroweak Model.     
     
The DL correction for QED processes dates back many years to the paper by     
Sudakov \cite{sudak}. In Section 2 we present the original Sudakov form     
factor for QED in a form in which it can be generalized to non-Abelian     
gauge theories, a subject to which we then turn. We start with the processes     
where all the scalar products $p_{l}p_{j}$ of the momenta of participating particles     
are of the same order. In Section 3 we compute the DL corrections to     
processes governed by massless non-Abelian gauge theories, and then, in     
Section 4, we turn to the interesting case of broken gauge symmetries,     
discussing, in particular, the DL effects for Standard Model electroweak     
processes.     
     
In all these cases the DL terms can be resummed in exponential forms     
which are generalizations of    
\begin{equation}     
{\cal M}\;=\;{\cal M}_{\rm Born} \exp \left( -\frac{1}{2}%
\sum_{i=1}^{n}W_{i}(s,\mu ^{2})\right) .  \label{eq:a1}     
\end{equation}     
Here $W_{i}$ is the probability of the emission of a soft and     
quasi-collinear gauge boson from an external particle $i$, and the summation     
is over all $n$ external charged particles participating in the process. The     
virtual particles are subject to an infrared cut-off $\mu $ on their     
transverse momenta.     
     
We use an equation for the evolution of the scattering amplitudes as a function of the infrared     
cut-off $\mu $ of the transverse momenta of virtual particles. Our approach     
is based on a gauge invariant dispersive method which is a generalization of     
a powerful theorem on photon Bremsstrahlung due to Gribov \cite{vg}. His     
remarkable theorem, proved by dispersive methods, states that, due to gauge     
invariance, the region of applicability of well known formulas for     
accompanying photon Bremsstrahlung is considerably extended at high energies.     
In these formulas the amplitude for a process with the emission of a soft     
photon is expressed, in a factorized form, in terms of the amplitude for the     
non-radiative process, with the particles taken {\it on-mass-shell}. Since     
the amplitude is taken on-shell, {\it gauge invariance} is guaranteed. The     
evolution equation approach greatly simplifies the computation of DL effects     
for electroweak processes, which are mediated by broken symmetry with the     
photon having components in both the $SU$(2) and $U$(1) gauge groups.     
     
The evolution equations for the amplitude as a function of the infrared     
cut-off parameter $\mu$ are analogous to the Renormalization Group Equations     
(RGE). Basically we start from the domain of very large $\mu$, where the     
Born amplitude applies, and evolve down to small $\mu$, matching the     
solution at the mass thresholds bounding every new kinematic domain. It is     
therefore not surprising that the structure of the exponentiation of the DL     
effects has a clear physical interpretation.     
     
Of course for observable processes it is necessary to consider the DL     
corrections to both the non-radiative and the radiative processes. For     
radiative processes we again have exponentiation of the Sudakov DL corrections in a form   
similar to (\ref{eq:a1}). The dependence on the infrared     
cut-off parameter $\mu $ is canceled in the measurable semi-inclusive     
process, and is replaced by a dependence on the experimental acceptance cuts     
on the soft boson emissions. In Sections 3.1, 4.1 and 4.2 we address these     
issues.     
     
As well as DL corrections of the Sudakov type, there are DL contributions     
specific to forward or 
backward scattering, when a final particle is     
quasi-collinear to an initial particle and their energies are almost equal.     
This is the domain of Regge kinematics and is the subject of Section 5. Here     
we have the DL corrections related to the exchange of soft pairs of fermions     
or of bosons in the crossed channel. The infrared evolution equations are     
now non-linear in the amplitude\footnote{%
An analogy is now to the {\it non-linear} RGE for the coupling constant,     
whereas for the Bremsstrahlung processes an analogy is the {\it linear} RGE     
for the moments of structure functions.}, and in the $j$-plane     
representation are of Ricatti-type form. In simple cases the latter forms     
can be reduced to a Schr\"{o}dinger-type equation with a harmonic potential.     
This approach can also be applied to production processes in the multi-Regge     
kinematics. In this case the evolution equations are solved in a sequence of     
domains starting from the region where the Born amplitude applies and     
evolving down in the cut-off $\mu $ to the region where soft particles are     
emitted. The vector boson and fermion reggeization can be also easily     
verified in the DL approximation. Moreover, the infrared evolution equation     
allows the construction of scattering amplitudes with quasi-elastic     
unitarity.     
     
\section{Sudakov form factor in QED}     
     
The high energy asymptotics of electromagnetic processes was calculated many     
years ago within the framework of QED \cite{sudak}--\cite{l}. In particular     
the amplitude for $e^{+}e^{-}$ elastic scattering at a fixed angle ($s\sim     
|t|\sim |u|\gg m^{2}\gg \lambda ^{2}$, where $m$ is the electron and $%
\lambda $ a fictitious \footnote{$\lambda$ plays the role of the infrared     
cut-off. In physical cross sections the divergence in $\lambda$ of the     
elastic amplitude is canceled with the analogous divergences in processes     
with soft photon emissions.} photon mass) in the DL approximation has the     
form \cite{bk}     
\begin{equation}     
{\cal M}={\cal M}_{\rm Born}\; \Gamma ^{2}\left( \frac{s}{m^{2}},\frac{m^{2}}{%
\lambda ^{2}}\right) \,,     
\end{equation}     
where ${\cal M}_{\rm Born}$ is the Born amplitude for $e^{+}e^{-}$ scattering     
and $\Gamma $ is the Sudakov form factor. The DL approximation applies in     
the energy regime     
\begin{equation}     
\alpha \log ^{2}\frac{s}{m^{2}}\sim \alpha \log \frac{s}{m^{2}}\log \frac{%
m^{2}}{\lambda ^{2}}\sim 1,     
\end{equation}     
where the QED coupling $\alpha =e^{2}/4\pi \ll 1$. Thus each charged     
external particle effectively contributes $\sqrt{\Gamma }$ to the total     
amplitude. The Sudakov form factor appears in the elastic scattering of an     
electron off an external field \cite{sudak}. It is of the form:     
\begin{equation}     
\Gamma \left( \frac{s}{m^{2}},\frac{m^{2}}{\lambda ^{2}}\right) =\exp \left(     
-\frac{\alpha }{2\pi }R\left( \frac{s}{m^{2}},\frac{m^{2}}{\lambda ^{2}}%
\right) \right)     
\end{equation}     
To specify $R$ it is convenient to use the Sudakov parametrization of the     
momentum of the exchanged virtual photon :     
\begin{equation}     
k=vp_{1}+up_{2}+k_{\perp }\,,     
\end{equation}     
where $p_{1}$ and $p_{2}$ are the initial and final momenta of the scattered     
electron. $R(s)$ can then be written as the integral over $u$ and $v$:     
\begin{equation}     
R\left( \frac{s}{m^{2}},\frac{m^{2}}{\lambda ^{2}}\right)     
=\int_{0}^{1}du\int_{0}^{1}dv\,\left( \frac{1}{u+m^{2} \; v/s}\right)     
\left( \frac{1}{v+m^{2} \; u/s}\right) \,\,\theta (suv-\lambda ^{2})\,,     
\end{equation}     
where $s\sim |t|\sim 2p_{1}p_{2}$. The first two factors in the integrand     
correspond to the propagators of the virtual fermions which occur in the     
one-loop triangle Sudakov diagram. The $\theta $ - function appears as a     
result of the integration of the propagator of the photon over its     
transverse momentum $k_{\perp }$, noting that the main contribution comes     
from the region near the photon mass shell \cite{sudak}:      
\begin{equation}     
suv=\lambda ^{2}+{\mbox{\boldmath $k$}_{\perp }^{2}}\,.  \label{eq:kp}     
\end{equation}     
To DL accuracy (6) gives for $\lambda \ll m$:      
\begin{equation}     
R\left( \frac{s}{m^{2}},\frac{m^{2}}{\lambda ^{2}}\right) =\frac{1}{2}\ln     
^{2}\frac{s}{m^{2}}+\ln \frac{s}{m^{2}}\ln \frac{m^{2}}{\lambda ^{2}}\,,     
\end{equation}     
where the result comes equally from two different kinematical regions, $v\gg     
u$ and $u\gg v$. Therefore one can write $R=2r$.     
     
We can obtain physical insight by presenting the two equal contributions     
separately. In the first region, with $v\gg u$, the virtual photon is     
emitted along $p_{1}$ and the parameter $v$ is given by the ratio of     
energies of the photon and the initial electron. Here instead of $u$, it is     
convenient to use Eq. (\ref{eq:kp}) to replace it by the square of the     
transverse momentum component of the photon. Then integrating over $v$ and ${%
\mbox{\boldmath     
$k$}_{\perp }^{2}}$ gives      
\begin{equation}     
r\left( \frac{s}{m^{2}},\frac{m^{2}}{\lambda ^{2}}\right) =     
\int_{\lambda /\sqrt{s}}^{1}\frac{dv}{v}\int_{\lambda ^{2}}^{sv^{2}}\frac{d{%
\mbox{\boldmath     
$k$}_{\perp }^{2}}}{{\mbox{\boldmath $k$}_{\perp }^{2}}+m^{2}v^{2}}\simeq     
\int_{\lambda ^{2}}^{s}\frac{d{\mbox{\boldmath $k$}_{\perp }^{2}}}{{%
\mbox{\boldmath     
$k$}_{\perp }^{2}}}\int_{|{\scriptsize {\mbox{\boldmath $k$}}_{\perp }|/%
\sqrt{s}}}^{{\rm min} (|{\scriptsize {\mbox{\boldmath $k$}}_{\perp }|/m,\;1)}}%
\frac{dv}{v}     
\end{equation}     
in the DL approximation, which may be evaluated to give half of $R$. The     
quantity $r$ is proportional to the probability $w_{i}$ of the emission of a     
soft and almost collinear photon from an external particle with energy $%
\sqrt{s}$ and mass $m_{i}$, i.e.     
\begin{equation}     
w_{i}(s,\lambda ^{2})=\frac{\alpha }{\pi }\,r\left( \frac{s}{m_{i}^{2}},%
\frac{m_{i}^{2}}{\lambda ^{2}}\right) .  \label{eq:a10}     
\end{equation}     
If several charged particles participate in a process, for example $%
e^{+}e^{-}\rightarrow f\bar{f}f\bar{f}$, then analogous contributions appear     
for each external line, provided all external invariants are large and of     
the same order. This leads to the general result     
\begin{equation}     
{\cal M}={\cal M}_{{\rm Born}}\exp     
\left( -\frac{1}{2}\sum_{i=1}^{n}w_{i}(s,%
\lambda^{2})\right) ,  \label{eq:a11}     
\end{equation}     
where $n$ is the number of external lines corresponding to charged     
particles. In summary the soft emissions described by the Sudakov form     
factor is a quasi-classical effect which does not depend on the hard     
dynamics of the process. In particular there are no quantum mechanical     
interference effects in the DL Sudakov corrections, for large scattering     
angles.     
     
\section{Generalization to non-Abelian gauge theories}     
     
Sudakov effects have been widely discussed for non-Abelian gauge theories,     
such as $SU(N)$ and can be calculated in various ways (see, for instance,     
\cite{NA}). We consider here the scattering amplitude in the simplest     
kinematics when all its invariants $s_{lj}=2p_{l}p_{j}$ are large and of the     
same order $s_{lj}\sim s$. A general method of finding the DL asymptotics     
(not only of the Sudakov type) is based on the infrared evolution equations     
describing the dependence of the amplitudes on the infrared cutoff $\mu $ of     
the virtual particle transverse momenta \cite{kl}. This cutoff plays the     
same role as $\lambda $ in QED, but, unlike $\lambda $, it is not     
necessary that it vanishes and it may take an arbitrary value. It can be introduced in     
a gauge invariant way by working, for instance, in a finite phase space     
volume in the transverse direction with linear size $l\sim 1/\mu $. Instead     
of calculating asymptotics of particular Feynman diagrams and summing these     
asymptotics for a process with $n$ external lines it is convenient to     
extract the virtual particle with the smallest value of $|{%
\mbox{\boldmath     
$k$}_{\perp }}|$ in such a way, that the transverse momenta $|{%
\mbox{\boldmath $k$}_{\perp }^{\prime }}|$ of the other virtual particles     
are much bigger     
\begin{equation}     
\mbox{\boldmath $k$}_{\perp }^{\prime ^{2}}\gg {\mbox{\boldmath $k$}_{\perp     
}^{2}}\gg \mu ^{2}\;.     
\end{equation}     
For the other particles ${\mbox{\boldmath $k$}_{\perp }^{2}}$ plays the role of     
the initial infrared cut-off $\mu ^{2}$.     
     
In particular, the Sudakov DL corrections are related to the exchange of     
soft gauge bosons, see Fig.~1. For this case the integral over the 
momentum $k$ of the     
soft (i.e. $|k^0|\ll \sqrt{s}$) virtual boson with the smallest 
${\mbox{\boldmath $k$}}_\perp$ can be factored off, which leads to the 
following infrared evolution equation:     
\begin{eqnarray}     
{\cal M}(p_1,...,p_n;\mu^2) & = & {\cal M}_{\rm Born}(p_1,...,p_n) -\frac{i}{2}     
\frac{g^2}{(2\pi)^4} \sum_{j,l=1, j \neq l}^n \int_{s \gg \mbox{{\scriptsize \boldmath $k$}}^2_\perp   
\gg \mu^2} \frac{d^4k}{k^2+i \epsilon} \;\;     
\frac{p_jp_l}{(kp_j)(kp_l)}  \nonumber \\     
& & \times \; T^a(j) T^a(l) {\cal M} (p_1,...,p_n;{\mbox{\boldmath $k$}%
^2_\perp}) \,,  \label{eq:vem}     
\end{eqnarray}     
where the amplitude ${\cal M}(p_1,...,p_n;{\mbox{\boldmath $k$}^2_\perp})$     
on the right hand side is to be taken on the mass shell, but with the     
substituted infrared cutoff: $\mu^2 \longrightarrow {\mbox{\boldmath $k$}%
^2_\perp}$. The generator $T^a(l) (a=1,...,N)$ acts on the color indices of     
the particle with momentum $p_l$. The non-Abelian gauge coupling is $g$.     
In Eq. (\ref{eq:vem}), and below, $\mbox{\boldmath $k$}_\perp$ denotes the component
of the gauge boson momentum $\mbox{\boldmath $k$}$ transverse to the particle 
emitting this boson. Note that in Sudakov DL corrections there are no interference
effects, so that we can talk about the emission (and absorption) of a gauge
boson by a definite (external) particle, namely by a particle with momentum
almost collinear to $\mbox{\boldmath $k$}$. It can be expressed in invariant form
as $\mbox{\boldmath $k$}^2_\perp \equiv \min ( (kp_l)(kp_j)/(p_lp_j))$ for all $j \neq l$.
\begin{figure} 
\centering 
\epsfig{file=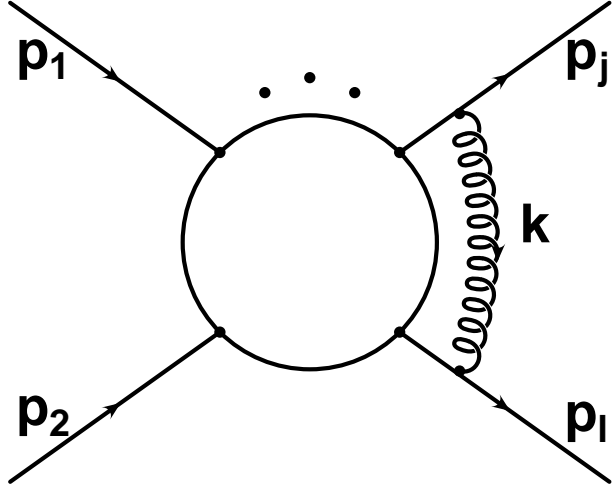,width=8cm} 
\caption{Feynman diagrams contributing to the infrared evolution 
equation (\ref{eq:vem}) for a process with $n$ external legs. In a general
covariant gauge the
virtual gluon with the smallest value of ${\mbox{\boldmath $k$}}_{\perp}$ is attached to 
different external lines. The inner scattering amplitude is assumed to be 
on the mass shell.}
\end{figure} 
The above factorization is related to a non-Abelian generalization of the     
Gribov theorem\footnote{The non-Abelian
generalization of Gribov's theorem is given in (\ref{eq:rem}) below, together with a description of 
its essential content.} for the amplitude of the Bremsstrahlung of a photon with     
small transverse momentum ${\mbox{\boldmath $k$}_{\perp }}$ in high energy hadron scattering     
\cite{vg}.  
 
The form in which we present
Eq. (\ref{eq:vem}) corresponds to a covariant gauge 
for the gluon with 
momentum $k$. Formally this expression can be written in a gauge invariant 
way if we include in the sum the term with $j=l$ (which does not give 
a DL contribution). Indeed, in this case we can 
substitute $p_{i}p_{j}$ by $-p_{i}^{\mu }p_{j}^{\nu }d_{\mu \nu }(k)$, where 
the polarization matrices of the boson $d_{\mu \nu }(k)$ in the various     
gauges differ by the terms proportional to $k^{\mu }$ or $k^{\nu }$ giving a
vanishing contribution due to the conservation of the total color charge $
\sum_{a}T^{a}=0$. Thus we have the possibility of choosing appropriate     
gauges for each kinematical region of quasi-collinearity of $k$ and     
$p_{l}$. We can, however, use (\ref{eq:vem}) as well, noting that in this region
for $j\neq l$ we have $p_{j}p_{l}/kp_{j}\simeq E_{l}/\omega $, where $E_{l}$
is the energy of the particle with momentum $p_{l}$ and $\omega$ the
frequency of the emitted gauge boson, so that:     
\begin{eqnarray}
{\cal M}(p_{1},...,p_{n};\mu ^{2}) &=&{\cal M}_{\rm Born}(p_{1},...,p_{n})-\frac{%
2g^{2}}{(4\pi )^{2}}\sum_{l=1}^{n}\int_{\mu ^{2}}^{s}\frac{d{%
\mbox{\boldmath $k$}_{\perp }^{2}}}{{\mbox{\boldmath $k$}_{\perp   
}^{2}}}\int_{|\mbox{\scriptsize \boldmath $k$}_{\perp }|/\sqrt{s}}^{{\rm min}   
(|\mbox{\scriptsize \boldmath $k$}_{\perp }|/m_l,\;1)}\frac{dv}{v}  \nonumber \\     
&&\times \;C_{l}{\cal M}(p_{1},...,p_{n};{\mbox{\boldmath     
$k$}_{\perp }^{2}}%
)\;\;,  \label{eq:dgvem}     
\end{eqnarray}     
where $C_{l}$ is the eigenvalue of the Casimir operator $T^{a}(l)T^{a}(l)$     
($%
C_{l}=C_{A}$ for gauge bosons in the adjoint representation of the gauge     
group $SU(N)$ and $C_{l}=C_{F}$ for fermions in the fundamental     
representation).     
     
The differential form of the infrared evolution equation follows immediately     
from (\ref{eq:dgvem}):     
\begin{equation}     
\frac{\partial {\cal M}(p_{1},...,p_{n};\mu ^{2})}{\partial \log (\mu     
^{2})}%
=K(\mu ^{2}){\cal M}(p_{1},...,p_{n};\mu ^{2})\,,  \label{eq:ee}     
\end{equation}     
where     
\begin{equation}     
K(\mu ^{2})\equiv -\frac{1}{2}\sum_{l=1}^{n}\frac{\partial W_{l}(s,\mu     
^{2})%
}{\partial \log (\mu ^{2})}     
\end{equation}     
with     
\begin{equation}     
W_{l}(s,\mu ^{2})=\frac{g^{2}}{4\pi^{2}}C_{l}\,r\left( \frac{s}{m_{l}^{2}},     
\frac{m_{l}^{2}}{\mu ^{2}}\right) \,.  \label{eq:wl}     
\end{equation}     
As in the Abelian case, $W_l$ is the probability to emit a soft and     
almost collinear gauge boson from the particle $l$ with mass     
$m_l$, subject to the infrared cut-off $\mu $ on the transverse momentum. Note  
again that the cut-off $\mu $ is not taken to zero. The function $r$     
is determined by (9) for arbitrary values of the ratio $m_l/\mu $. To     
logarithmic accuracy, we obtain from (\ref{eq:wl}):     
\begin{equation}     
\frac{\partial W_{l}(s,\mu ^{2})}{\partial \log (\mu ^{2})}=-\frac{g^{2}}{%
8\pi ^{2}}C_{l}\log \frac{s}{\max (\mu ^{2},m_{l}^{2})}\,.     
\end{equation}     
The infrared evolution equation (\ref{eq:ee}) should be solved with an     
appropriate initial condition. In the case of large scattering angles, if we    
choose the cut-off to be the large scale $s$ then clearly there are no     
Sudakov corrections. The initial condition is therefore     
\begin{equation}     
{\cal M}(p_{1},...,p_{n};s)={\cal M}_{\rm Born}(p_{1},...,p_{n}),     
\end{equation}     
and the solution of (\ref{eq:ee}) is thus given by the product of the Born     
amplitude and the Sudakov form factors:     
\begin{equation}     
{\cal M}(p_{1},...,p_{n};\mu ^{2})={\cal M}_{\rm Born}(p_{1},...,p_{n})\exp     
\left( -\frac{1}{2}\sum_{l=1}^{n}W_{l}(s,\mu ^{2})\right)     
\end{equation}     
Therefore we obtain an exactly analogous Sudakov exponentiation for the     
gauge group $SU(N)$ to that for the Abelian case, see (\ref{eq:a11}).     
Theories with semi-simple gauge groups can be considered in a similar way.     
     
\subsection{DL corrections to processes with soft emissions}     
     
Since ultimately we are interested in measurable cross sections we have to     
consider the DL corrections to amplitudes of processes with soft emissions,     
as well as those without. Only in inclusive cross sections will the     
dependence on the infrared cut-off $\mu^2$ disappear, being replaced by     
parameters specifying the experimental acceptance. To put it in another way,     
cross sections of the emission processes receive large (DL) contributions     
from regions where the emitted bosons are soft and the emission angles are     
small. Therefore, to be consistent, we need to calculate cross sections of     
such processes as well. This is easy to do for QED processes, where the     
single gauge boson (the photon) is neutral and does not possess     
self-interactions. Therefore soft photons are emitted independently     
according to a Poisson distribution. In non-Abelian theories, gauge bosons     
are not neutral and interact with each other. Consequently, the soft     
emission does not follow a Poisson distribution \cite{kf}.     
 
We again consider the simplest situation, when the additional     
soft gauge boson is emitted in the process with all invariants $s_{lj}$     
large. Of course, for the emission of a boson almost collinear to the     
particle the direction of the particle with momentum $p_{i}$, the invariant     
$%
2kp_{i}$ is small in comparison with $s$. In the case of non-Abelian gauge     
theories the corresponding amplitude for the emission of a soft gauge boson     
with small ${\mbox{\boldmath $k$}_{\perp }^{2}}\ll \mu ^{2}$ has, according     
to the Gribov theorem, the following form:     
\begin{equation}     
{\cal M}^{a}(p_{1},...,p_{n};k;\mu ^{2})=\sum_{j=1}^{n}g\;\frac{\varepsilon     
^{*}p_{j}}{kp_{j}}\; 
T^{a}(j)\;{\cal M}(p_{1},...,p_{n};\mu ^{2})\,.     
\label{eq:rem}     
\end{equation}     
The possible corrections to this factorized expression are of the order of     
${%
\mbox{\boldmath $k$}_{\perp }^{2}}/\mu ^{2}$. However, to DL accuracy,     
we can substitute $\mu ^{2}$ in the arguments of the scattering amplitudes     
by its boundary value ${\mbox{\boldmath $k$}_{\perp }^{2}}$. Notice that the     
amplitude on the r.h.s. of (\ref{eq:rem}) is taken on-the-mass shell, which     
guarantees its gauge invariance. The result (\ref{eq:rem}) is highly     
non-trivial in the Feynman diagram approach. It means, that the region of     
applicability of the classical formulas for the Bremsstrahlung amplitudes is     
significantly enlarged at high energies. V.~Gribov proved this result by     
using dispersion relations in the variables $(k+p_{j})^{2}$, and     
demonstrating that for small $|{\mbox{\boldmath $k$}_{\perp }}|$ the pole     
terms in these invariants are much larger than the corresponding cut     
contributions due to the gauge invariance of the theory \cite{vg}. The     
region of applicability of (\ref{eq:rem}) corresponds to the situation when     
the momentum of the emitted soft boson does not spoil the kinematics of the     
non-radiative process. This implies that the frequency $\omega $ of the     
boson emitted off the particle with momentum $p_{l}$ should be much smaller     
than the energy $E_{l}$, and that the emission angle $\vartheta _{l}\simeq     
|{%
\mbox{\boldmath $k$}_{\perp }}|/\omega$ should be much smaller than the angle     
between $p_{l}$ and all other momenta $p_{j}$ (otherwise one needs to     
include interference effects). For the physical Coulomb gauge ($\varepsilon     
_{0}=0,\,\,\mbox{\boldmath $\varepsilon$}\cdot \mbox{\boldmath $k$}=0$) in     
the kinematical region where the gauge boson is emitted at a small angle $%
\vartheta _{i}$ with respect to the particle with momentum $p_{i}$ in (\ref     
{eq:rem}) only the term with $j=i$ contributes.     
\begin{figure} 
\centering 
\epsfig{file=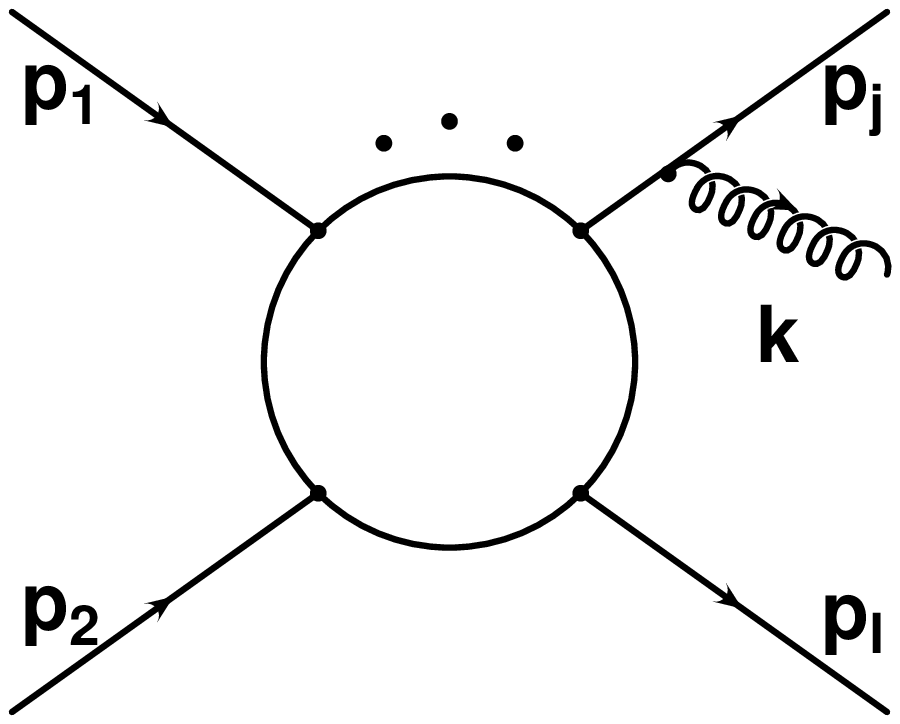,width=6cm} 
\hspace{1cm} 
\epsfig{file=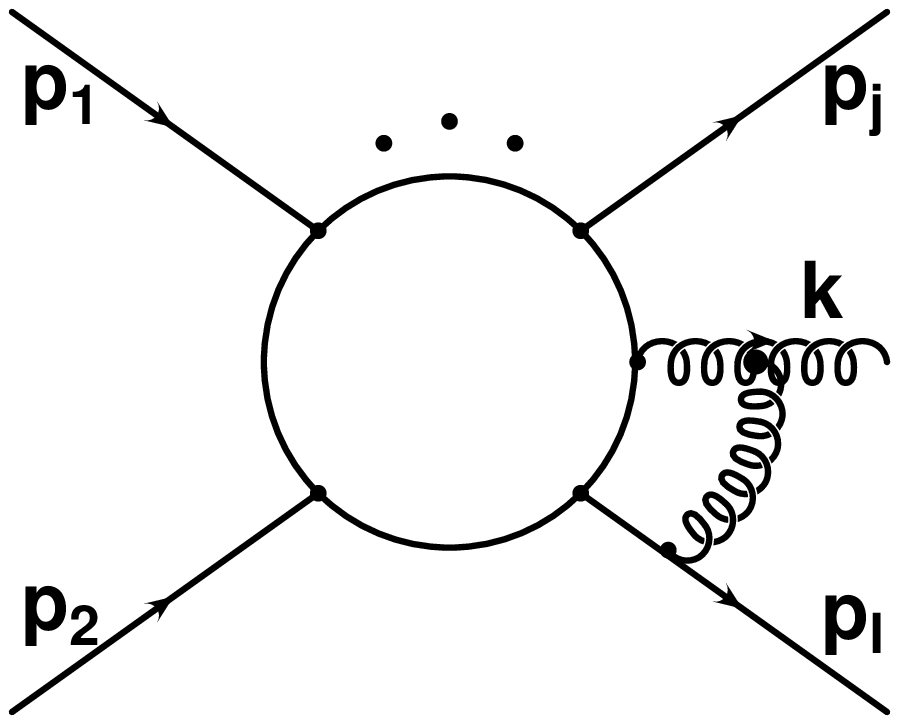,width=6cm} 
\caption{Schematic Feynman diagrams contributing to the real gauge 
boson emission for a process with $n$ external legs.  For $\mbox{\boldmath $k$}_\perp^2 \ll \mu^2$ 
the diagram on the left corresponds to 
the non-Abelian generalization of Gribov's theorem (\ref{eq:rem}).  
The diagram on the right leads 
to an additional term, (\ref{eq:f1}), in the kernel of the evolution equation in the case when 
$\mbox{\boldmath $k$}_\perp^2 \gg \mu^2$.} 
\end{figure} 
The method based on the infrared evolution equation allows us to calculate,     
in the DL approximation, the amplitudes of the hard processes accompanied by    
the emission of any number of soft gauge bosons \cite{kf}. Let us consider     
the amplitude for the emission of one soft gauge boson in the region of     
quasi-collinearity of its momentum with momentum $p_{i}$ (i.e., the emission    
of a soft boson by a particle with momentum $p_{i}$). When the transverse     
momentum $|{\mbox{\boldmath $k$}_{\perp }}|$ of this boson is much less than    
the infrared cut-off $\mu $, used for the virtual particles, the amplitude     
is given by the term with $j=i$ in the sum in (\ref{eq:rem}). But we need to    
know the emission amplitude in the opposite case, $|{\mbox{\boldmath 
$k$}_{\perp }}|\gg \mu $. It can be found from the evolution equation in this   
region using expression (\ref{eq:rem}) at $\mu =|{\mbox{\boldmath 
$k$}_{\perp }}|$ as the initial condition. The kernel of the evolution equation 
in this region differs from the corresponding kernel in the region $|{%
\mbox{\boldmath $k$}_{\perp }}|\ll \mu $ (that is the kernel (16) of the     
evolution equation for the amplitude without emission) by a term connected     
with the emission of a virtual boson from the real gauge boson with momentum    
$k$:     
\begin{equation}     
\Delta K(\mu ^{2})=-\frac{1}{2}\frac{\partial W_{A}({\mbox{\boldmath $k$}%
_{\perp }^{2}},\mu ^{2})}{\partial \log (\mu     
^{2})}=\frac{g^{2}}{(4\pi )^{2}}%
C_{A}\log \frac{{\mbox{\boldmath $k$}_{\perp }^{2}}}{\mu ^{2}}\,,     
\label{eq:f1}     
\end{equation}     
where $W_{A}$ is given by (\ref{eq:wl}) with $C_{l}=C_{A}$ and $m_{l}=\mu$, see Fig.~2.     
It is clear, that this new term in the kernel for evolution from $%
\mbox{\boldmath $k$}_{\perp }^{2}$ to $\mu ^{2}$ leads to an additional term     
$W_{A}({\mbox{\boldmath     
$k$}_{\perp }^{2}},\mu ^{2})$ in the Sudakov exponential. Thus, the     
amplitude for the emission of one gauge boson with small transverse momentum     
from the hard scattering process is of the form:     
\[     
{\cal M}^{a}(p_{1},...,p_{n};k;\mu ^{2})=\sum_{j=1}^{n}g\;\frac{\varepsilon     
^{*}p_{j}}{kp_{j}}\;T^{a}(j)\;{\cal M}_{\rm Born}(p_{1},...,p_{n})     
\]     
\begin{equation}     
\times \exp \left( -\frac{1}{2}\sum_{l=1}^{n}W_{l}(s,\mu ^{2})-\frac{1}{2}%
W_{A}({\mbox{\boldmath $k$}_{\perp }^{2}},\mu ^{2})\right) \,.     
\label{eq:f2}     
\end{equation}     
We note again that here
$\mbox{\boldmath $k$}_{\perp }^{2}$ means the square of the component     
of the three-dimensional momentum transverse to the momentum of the emitting     
particle, say for example, $p_{l}$. We can write $\mbox{\boldmath $k$}%
_{\perp }^{2}$ in the invariant form $(kp_{l})(kp_{j})/(p_{l}p_{j})$ with $%
j\neq l$, which does not depend on $j$.     
     
Again we see that we have the exponentiation of the Sudakov DL corrections. Note that     
in the Abelian QED case we have $W_A=0$ and the exponent for the photon     
Bremsstrahlung amplitude remains the same as that for the process without     
photon emission. It is related to the Poisson distribution for soft photon     
production.     
     
The exponentiation of virtual DL corrections holds for multiple emission     
processes as well. In QED it is trivial, since the soft photons are emitted     
independently. In non-Abelian gauge theories it is not so simple. The main     
complexity is connected with the nontrivial structure of the amplitudes for     
multiple emission of real soft gauge bosons, arising from their     
self-interaction. But in the DL approximation these amplitudes can be     
calculated. Due to the coherence effect, the branching cascade develops only     
in the region of sequentially shrinking angular cones \cite{kf}. In this     
region the Born amplitudes for multiple emission processes have a factorized form     
and the virtual corrections exponentiate \cite{kf}. It is proved by solving the     
infrared evolution equation in a series of regions where the infrared     
cut-off $\mu $ is bounded between a sequence of decreasing transverse     
momenta of the emitted gluons, using in each region the solution of the     
previous region as the initial condition. The final result is     
     
\[     
{\cal M}(p_{1},...,p_{n};k_{1},...,k_{r};\mu ^{2})={\cal M}%
_{\rm Born}(p_{1},...,p_{n};k_{1},...,k_{r})     
\]     
\begin{equation}     
\exp \left( -\frac{1}{2}\sum_{l=1}^{n}W_{l}(s,\mu ^{2})-\frac{1}{2}%
\sum_{i=1}^{r}W_{A}({\mbox{\boldmath $k$}_{i\perp }^{2}},\mu ^{2})\right)     
\,,     
\label{eq:f3}     
\end{equation}     
where $k_{i}$ are the momenta of the emitted gluons with strongly ordered     
energies and ${\mbox{\boldmath $k$}_{i\perp }}$ are their components     
transverse to the momenta of the emitting jets.     
     
\section{Sudakov effects in broken gauge theories}     
     
The same method, based on the infrared evolution equation, is also     
applicable to broken gauge theories. Let us consider for definiteness the     
Standard electroweak theory, where the physical gauge bosons are a massless     
photon (described by the field $A_{\nu }$) and massive $W^{\pm }$ and $Z$     
bosons (described correspondingly by fields $W_{\nu }^{\pm }$ and     
$Z_{\nu }$%
). To DL accuracy, all masses can be set equal:     
\[     
M_{Z}\sim M_{W}\sim M_{\rm Higgs}\sim M     
\]     
and the energy considered to be much larger, $\sqrt{s}\gg M$. In the     
unbroken phase the corresponding Abelian and Yang-Mills fields are denoted     
by $B_{\nu }$ and $W_{\nu }^{a}$, with $a=1,2,3$. The physical fields are     
linear combinations of the fields 
 in the unbroken theory with coefficients     
depending on the Weinberg angle $\theta _{w}$. The left and right handed     
fermions are correspondingly doublets ($T=1/2$) and singlets ($T=0$) of the     
$%
SU$(2) weak isospin group and have hypercharge $Y$ related to the electric     
charge $Q$, measured in units of the proton charge, by the     
Gell-Mann-Nishijima formula $Q=T^{3}+Y/2$.     
     
In the evolution equation in the DL approximation the value of the infrared     
cutoff $\mu$ can be chosen in two different ranges : 1) $\sqrt{s} \gg \mu     
\gg M$ and 2) $\mu \ll M$. The second case is universal in the sense that it     
does not depend on details of the electroweak theory. It will be discussed     
below. In the first region we can neglect spontaneous symmetry breaking     
effects, in particular gauge boson masses, and consider the evolution     
equation in the unbroken phase with effectively massless particles $B$ and     
$%
W^a$. Of course one could calculate everything also in terms of the physical     
fields $A^\nu, Z^\nu$ and $W_\nu^\pm$. In the unbroken phase this is     
equivalent to the description in terms of the original fields $B^\nu$ and $%
W^a_\nu$ and leads to the same final result, but the intermediate steps will     
be more complicated because there are cancellations between     
non-exponentiating terms from diagrams with $Z$ and $\gamma$ exchanges. The     
separation of these contributions is not gauge invariant and if we would     
consider the diagrams without virtual photons we would violate $SU(2)\times     
U(1)$ symmetry. Taking into account only such diagrams leads to     
nonexponentiating DL effects in an axial gauge \cite{Kuhn}.  
Fig.~3 illustrates this at the two loop level. The loss     
of gauge invariance is related to the fact that the photon field contains     
the component $W_{\nu}^3$ of the non-Abelian field $W^a_{\nu}$, and so     
omitting the virtual photons would violate the conservation of the weak     
isospin current (in the unbroken theory).     
\begin{figure} \label{fig:zg} \vspace{-0.7cm}
\centering 
\epsfig{file=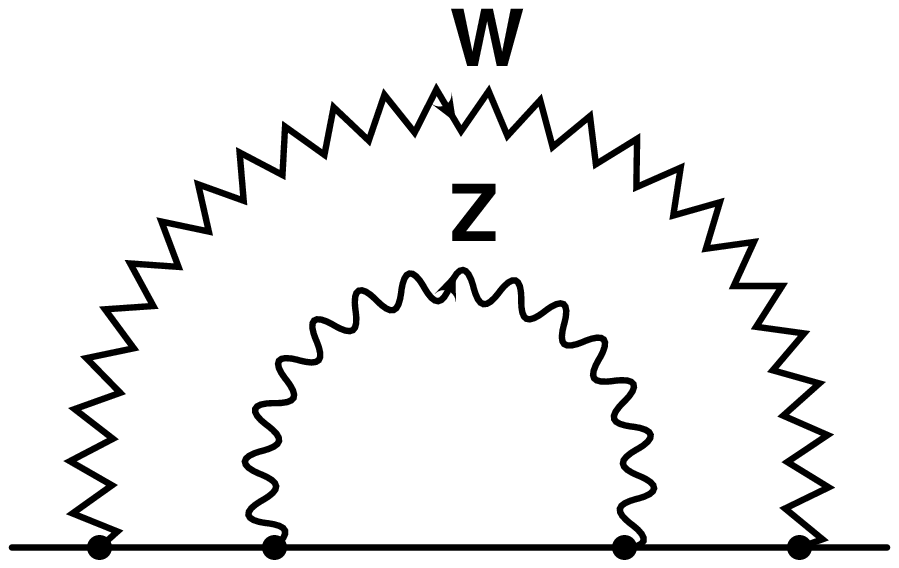,width=6cm} 
\hspace{1cm} 
\epsfig{file=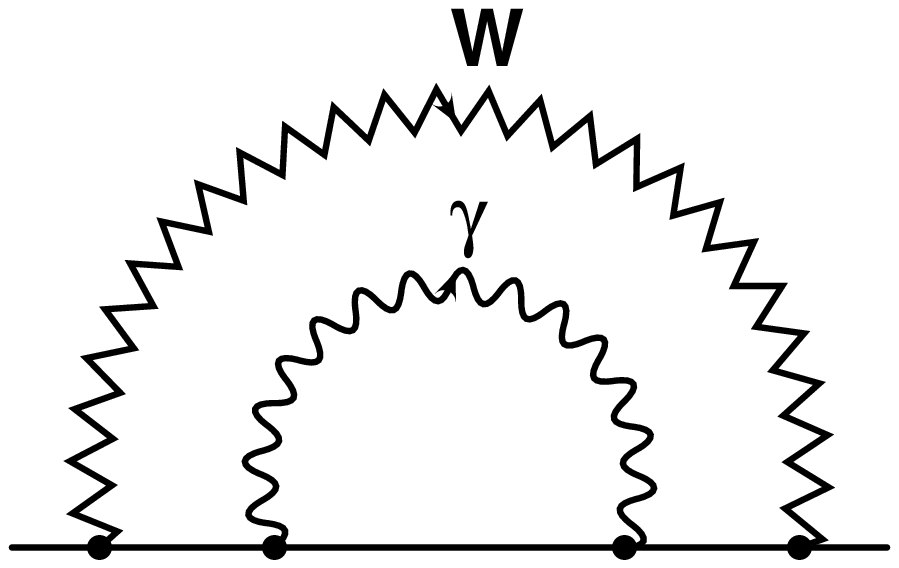,width=6cm} \\ 
\vspace{0.1cm}
\epsfig{file=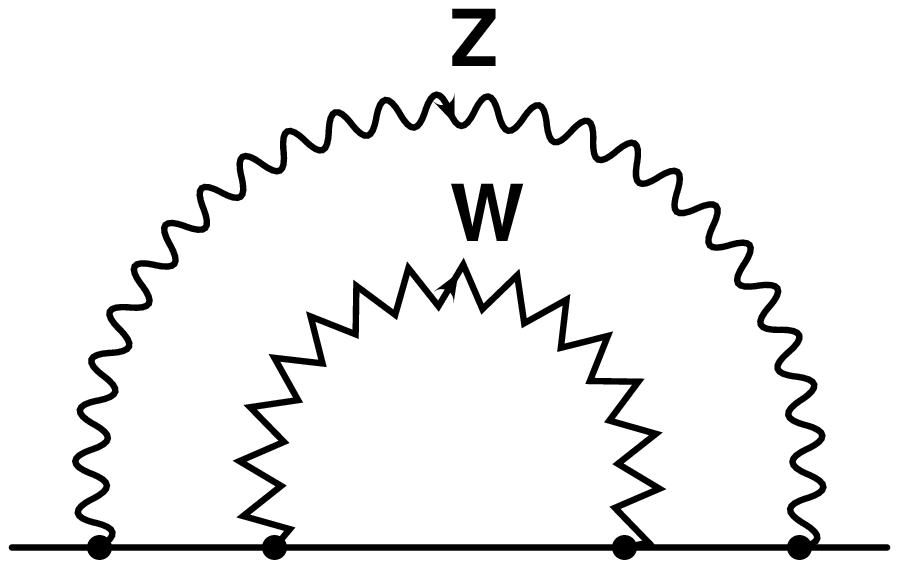,width=6cm} 
\hspace{1cm} 
\epsfig{file=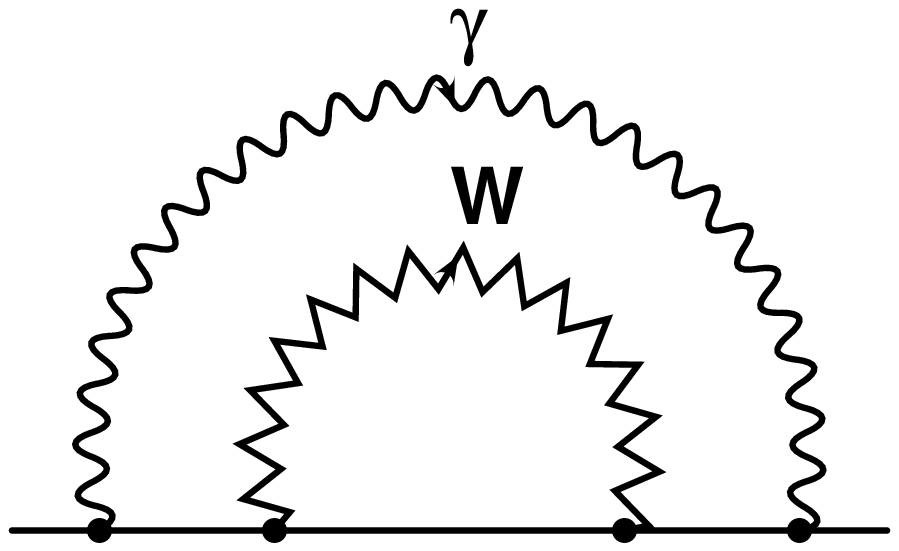,width=6cm} \\
\vspace{0.1cm}
\epsfig{file=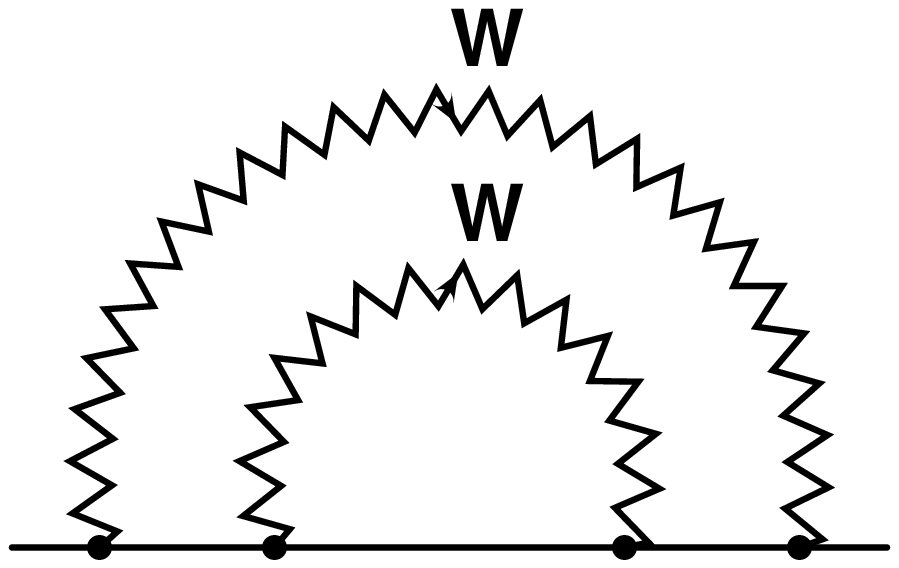,width=6cm}  
\hspace{1cm} 
\epsfig{file=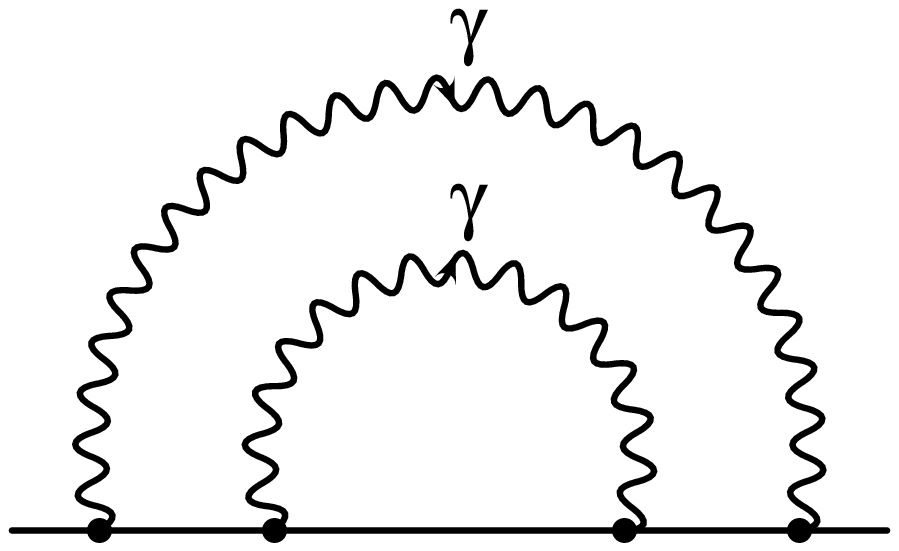,width=6cm} \\
\vspace{0.1cm}
\epsfig{file=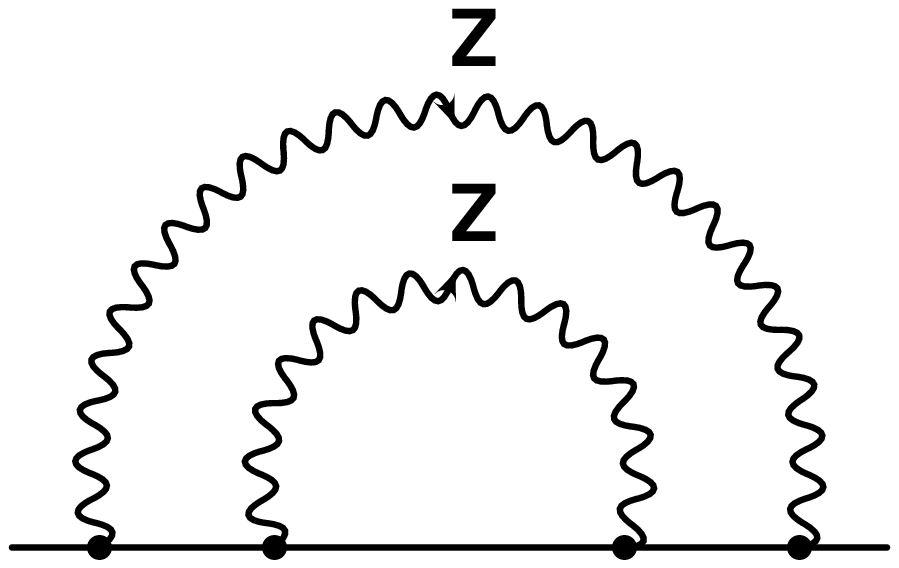,width=6cm}  
\hspace{1cm} 
\epsfig{file=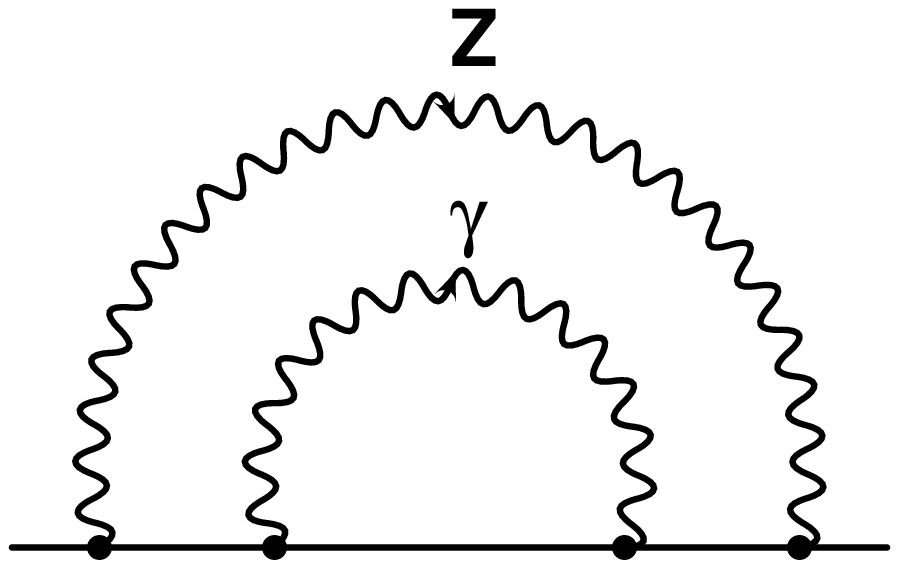,width=6cm} \\
\vspace{0.1cm}
\hspace{7cm} 
\epsfig{file=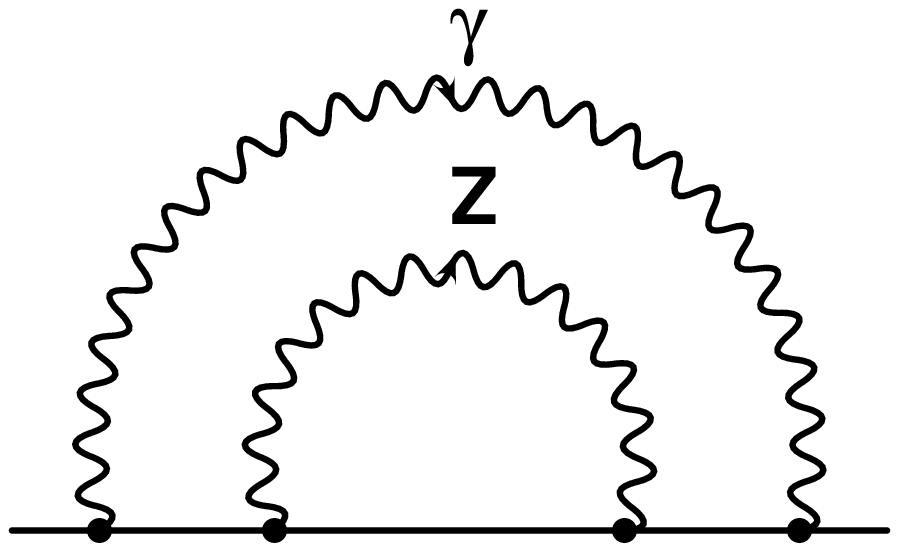,width=6cm}  
\caption{Two-loop `rainbow' Feynman diagrams contributing to 
DL corrections in an axial gauge. 
The photon contribution 
has DL corrections in both the regions $\mu^2 \ll M^2$ and
$M^2 \ll \mu^2 \ll s$. Taken together with the 
$W$- and $Z$-contributions, it
yields the exponentiation of the Sudakov DL terms in the electroweak theory.
In the region $M^2 \ll \mu^2$, 
the spontaneously broken gauge
symmetry is restored and omitting the photon contributions would
lead to a non-gauge invariant result.} 
\end{figure} 
In region 1) the infrared evolution equation, written in terms of the     
unbroken fields, is of a form analogous to (\ref{eq:ee}) if we assume, for     
simplicity, that all the charged particles have masses $m_{i}\leq M$%
\begin{equation}     
\frac{\partial {\cal M}(p_{1},...,p_{n};\mu ^{2})}{\partial \log (\mu     
^{2})}=%
\frac{\log (s/\mu ^{2})}{(4\pi )^{2}}\sum_{i=1}^{n}\left(     
g^{2}T_{i}(T_{i}+1)+{g^{\prime }}^{2}\left( \frac{Y_{i}}{2}\right)     
^{2}\right) {\cal M}(p_{1},...,p_{n};\mu ^{2})\,.  \label{eq:ubee}     
\end{equation}     
Here $T_{i}$ is the total weak isospin of particle $i$, $Y_{i}$ is its     
weak hypercharge, and 
$g$ and $g^{\prime }=g\tan \theta _{w}$ are the couplings     
of the $SU(2)$ and $U(1)$ gauge groups, respectively. The sum in (\ref     
{eq:ubee}) is to be performed over all $n$ external particles. As before,     
the initial condition is given by the requirement that for the
infrared cut-off $\mu     
^{2}=s$ we obtain the Born amplitude. The solution of (\ref{eq:ubee}) is     
thus given by     
\begin{eqnarray}     
{\cal M}(p_{1},...,p_{n};\mu ^{2}) &=&{\cal M}_{\rm Born}(p_{1},...,p_{n})     
\nonumber \\     
&&\times \exp \left[ -\frac{\log ^{2}(s/\mu ^{2})}{2(4\pi )^{2}}%
\sum_{i=1}^{n}\left( g^{2}T_{i}(T_{i}+1)+{g^{\prime }}^{2}\left( \frac{Y_{i}     
}{2}\right) ^{2}\right) \right] .  \label{eq:ubs}     
\end{eqnarray}     
The expression in the brackets in the exponential can be written in terms of     
the parameters of the broken theory as follows:     
\[     
g^{2}T_i(T_i+1)+{g^{\prime }}^{2}\left( \frac{Y_i}{2}\right)     
^{2}=e_i^{2}+g^{2}\left( T_i(T_i+1)-(T_i^{3})^{2}\right) +\frac{g^{2}}{\cos     
^{2}\theta _{w}}\left( T_i^{3}-\sin ^{2}\theta _{w}Q_i\right) ^{2},     
\]     
where the three terms on the r.h.s. correspond to the contributions of the     
soft photon (interacting with the electric charge $e_i=Q_ig\sin \theta _{w}$),     
the $W^{\pm }$ and the $Z$ bosons, respectively. Although we may rewrite     
solution (\ref{eq:ubs}) in terms of the parameters of the broken theory in     
the form of a product of three exponents corresponding to the exchanges of     
photons, $W^{\pm }$ and $Z$ bosons, it would be wrong to identify the     
contributions of the diagrams without virtual photons with this expression     
for the particular case $e_i^{2}=0$. This becomes evident when we note that if     
we were to omit photon lines then the result would depend on the choice of     
gauge, and therefore be unphysical. Only for $\theta _{w}=0$, where the     
photon coincides with the $B$ gauge boson, would the identification of the     
$%
e_i^{2}$ term with the contribution of the diagrams with photons be correct.     
     
Choosing the cutoff $\mu$ in the second region, $\mu \ll M$, the infrared     
evolution equation takes the following form:     
\begin{equation}     
\frac{\partial {\cal M}(p_{1},...,p_{n};\mu ^{2})}{\partial \log (\mu     
^{2})}%
=\sum_{i=1}^{n}\frac{e_{i}^{2}}{(4\pi )^{2}}\;\log \left( \frac{s}{\max     
(m_{i}^{2},\mu ^{2})}\right) {\cal M}(p_{1},...,p_{n};\mu ^{2}).     
\label{eq:ubeea}     
\end{equation}     
Evidently, only the photon contribution remains in this region. Now the     
appropriate initial condition is given by (\ref{eq:ubs}) evaluated at the     
matching point $\mu =M$. The solution is thus     
\begin{eqnarray}     
{\cal M}(p_{1},...,p_{n};\mu ^{2}) &=&{\cal M}_{\rm Born}(p_{1},...,p_{n})     
\nonumber \\     
&&\times \exp \left[ -\frac{g^{2}}{2(4\pi )^{2}}\sum_{i=1}^{n}\left( \left\{     
T_{i}(T_{i}+1)+\tan ^{2}\theta _{w}\left( \frac{Y_{i}}{2}\right)     
^{2}\right\} \log ^{2}\frac{s}{M^{2}}\right) \right]  \nonumber \\     
&&\times \exp \left[ - \frac{1}{2} \sum_{i=1}^{n} Q^2_i \left( w_i(s,\mu^2) 
- w_i(s,M^2) \right) \right] \nonumber \\ 
&=&{\cal M}_{\rm Born}(p_{1},...,p_{n})     
\nonumber \\     
&&\times \exp \left[ -\frac{g^{2}}{2(4\pi )^{2}}\sum_{i=1}^{n}\left( \left\{     
T_{i}(T_{i}+1)+\tan ^{2}\theta _{w}\left( \frac{Y_{i}}{2}\right)     
^{2}\right\} \log ^{2}\frac{s}{M^{2}}\right) \right]  \nonumber \\     
&&\times \exp \left[ -\sum_{i=1}^{n} \! \left( \frac{e_{i}^{2}}{(4\pi )^{2}}%
\! \left( \log \frac{s}{m_{i}\,M}\log \frac{M^{2}}{m_{i}^{2}}+\log \frac{s}{%
m_{i}^{2}}\log \frac{m_{i}^{2}}{\mu ^{2}} \! \right) \! \right) \! \right] ,     
\label{eq:ubsa}     
\end{eqnarray}     
where the last equality holds for $\mu \ll m_i$ and 
$m_i^2 \ll M^2$ from the respective 
expansions of $w_i$ in (\ref{eq:a10}). 
Let us stress that (\ref{eq:ubs}) and (\ref{eq:ubsa}) are applicable for     
processes involving chiral fermions as well as gauge bosons, provided that     
all the invariants are large (${\cal O} ( s )$) compared to $M^2$. Note,     
that in the case, when quarks or gluons participate in the reaction, we     
should multiply these expressions by the Sudakov factors corresponding to     
the virtual gluons emitted by these colored particles. The infrared     
evolution equations (\ref{eq:ubee}) and (\ref{eq:ubeea}) have a clear     
physical  
meaning analogous to the renormalization group equations and     
therefore it is natural to expect that the next-to-leading corrections to     
the kernels can be calculated.     
     
For physical observables soft real photon emission must be taken into     
account in an inclusive way and effectively the parameter $\mu^2$ in (\ref     
{eq:ubsa}) will be replaced by parameters depending on the experimental     
requirements.     
     
\subsection{DL effects for electroweak processes with soft emission}     
     
The calculation of amplitudes for processes with the emission of a gauge     
boson in the kinematical region which gives DL contributions to the cross     
sections (i.e. the region of soft quasi-collinear emission) is similar to     
the analogous calculation for unbroken gauge theories, with complications of     
the type that we discussed above. One is that we have to consider separately     
two regions of $|{\mbox{\boldmath $k$}_{\perp }}|$ of the emitted boson:     
first $|{\mbox{\boldmath     
$k$}_{\perp }}|\ll M$ and second $|{\mbox{\boldmath $k$}_{\perp }}|\gg M$. At     
high energies the cross sections of the emission processes receive DL     
contributions from both of these regions. Of course, $W^{\pm }$ and $Z$ boson     
emissions contribute in the second region only. Therefore consideration of     
the first case is very simple. For values of the infrared cut-off $\mu     
^{2}\gg {\mbox{\boldmath     
$k$}_{\perp }^{2}}$ the amplitude for the emission of a soft photon by a     
particle with momenta $p_{l}$ (i.e. emission within the cone along $%
\mbox{\boldmath $p_l$}$ not containing the momenta of the other particles)     
has, in the physical (Coulomb) gauge, a factorized form     
\begin{equation}     
{\cal M}(p_{1},...,p_{n};k;\mu ^{2})=e_{l}\;\frac{\varepsilon ^{*}p_{l}}{%
kp_{l}}\;{\cal M}(p_{1},...,p_{n};\mu ^{2})\,,  \label{eq:g1}     
\end{equation}     
according to the Gribov theorem of (\ref{eq:rem}). However, if ${%
\mbox{\boldmath     
$k$}_{\perp }^{2}}\ll M^{2}$, the kernel of the infrared evolution equation     
does not change when the cut-off $\mu ^{2}$ changes from the domain $\mu     
^{2}\gg {\mbox{\boldmath $k$}_{\perp }^{2}}$ to $\mu ^{2}\ll {%
\mbox{\boldmath     
$k$}_{\perp }^{2}}$ (since the $W^{\pm }$ and $Z$ bosons do not contribute in     
this first region). Therefore, (\ref{eq:g1}) remains valid at arbitrary     
values of the cut-off $\mu $.     
     
In the second region, $|\mbox{\boldmath $k$}_{\perp }|\gg M$, the result is     
more involved. We need to start from $\mu ^{2}\gg {\mbox{\boldmath $k$}%
_{\perp }^{2}}$, where we can use a generalization of Gribov's theorem.     
Consider again the case when a gauge boson is emitted by a particle with     
momentum $p_{l}$. We have     
\begin{equation}     
{\cal M}^{a}(p_{1},...,p_{n};k;\mu ^{2})=G_{0}^{a}(l)\frac{\varepsilon     
^{*}p_{l}}{kp_{l}}\;{\cal M}(p_{1},...,p_{n};\mu ^{2})\,,  \label{eq:g2}     
\end{equation}     
where ${\cal M}(p_{1},...,p_{n};\mu ^{2})$ is given by (\ref{eq:ubs}) and     
\begin{eqnarray}     
G_{0}^{\pm }=\frac{g}{\sqrt{2}}T^{\mp } &&{\rm for}\quad W^{\pm }\;{\rm %
emission},  \nonumber  \label{eq:a31} \\     
G_{0}^{Z}=\frac{g}{\cos \theta _{w}}\left( T^{3}-Q\sin ^{2}\theta     
_{w}\right) &&{\rm for}\quad Z\;{\rm emission},  \nonumber \\     
G_{0}^{\gamma }=Qg\sin \theta _{w} &&{\rm for}\quad \gamma \;{\rm emission},     
\end{eqnarray}     
with $Q=(T^{3}+Y/2)$. Then we have to solve the evolution equation in the     
region $M^{2}\ll \mu ^{2}\ll {\mbox{\boldmath $k$}_{\perp }^{2}}$ with the     
initial condition given by (\ref{eq:g2}) at the matching point $\mu ^{2}={%
\mbox{\boldmath $k$}_{\perp }^{2}}$. In fact, it is more appropriate to work     
in terms of the fields $W_{\nu }^{a}$ and $B_{\nu }$. The kernel of the     
evolution equation remains unchanged for the emission of the $B$-particle     
due to its Abelian nature. On the other hand the emission of the $W^{a}$     
-particle leads to the same additional contribution as in the unbroken     
theory (see (\ref{eq:f1})). Therefore in the cut-off region ${%
\mbox{\boldmath     
$k$}_{\perp }^{2}}\gg \mu ^{2}\gg M^{2}$, we obtain     
\begin{equation}     
{\cal M}^{a}(p_{1},...,p_{n};k;\mu ^{2})=G_{1}^{a}(l)\frac{\varepsilon     
^{*}p_{l}}{kp_{l}}\;{\cal M}(p_{1},...,p_{n};\mu ^{2})\,,    
\end{equation}     
with the same amplitude ${\cal M}(p_{1},...,p_{n};\mu ^{2})$ as in (\ref     
{eq:ubs}) and     
\begin{eqnarray}     
G_{1}^{\pm }&=&G_{0}^{\pm }\exp \left( -\frac{1}{2}W_{A}({\mbox{\boldmath     
$k$}%
_{\perp }^{2}},\mu ^{2})\right) \,, \nonumber \\    
G_{1}^{Z}&=&G_{0}^{Z}+g\cos \theta _{w}T^{3}\left( \exp \left( -\frac{1}{2}    
W_{A}(\mbox{\boldmath $k$}_{\perp }^{2},\mu ^{2}) \right) -1 \right) \,,     
\nonumber \\
G_{1}^{\gamma }&=&G_{0}^{\gamma }+g\sin \theta _{w}T^{3}\left( \exp \left( -%
\frac{1}{2}W_{A}({\mbox{\boldmath $k$}_{\perp }^{2}},\mu ^{2})\right)     
-1\right) \,.  \label{eq:g4}     
\end{eqnarray}     
Finally we study the region of the infrared cut-off $\mu ^{2}\ll M^{2}$. In     
this region the kernel of the evolution equation is determined by the     
electromagnetic interaction only; therefore, the only contribution related     
to the emitted particles is that for $W^{\pm }$emission. This contribution     
is given by     
\begin{equation}     
\Delta K_{W}(\mu ^{2})=-\frac{1}{2}\frac{\partial w_{W}({\mbox{\boldmath     
$k$}%
_{\perp }^{2}},\mu ^{2})}{\partial \log (\mu     
^{2})}=\frac{e^{2}}{(4\pi )^{2}}%
\log \frac{{\mbox{\boldmath $k$}_{\perp }^{2}}}{M^{2}}\,     
\end{equation}     
where $w_{W}({\mbox{\boldmath $k$}_{\perp }^{2}},\mu ^{2})$ is defined by   
(\ref{eq:a10}) and (9) with $M_{W}^{2}=M^{2}$. Consequently, for values of the     
infrared cut-off $\mu ^{2}\ll M^{2}$ we obtain:     
\begin{equation}     
{\cal M}^{a}(p_{1},...,p_{n};k;\mu ^{2})=G^{a}(l)\frac{\varepsilon     
^{*}p_{l}%
}{kp_{l}}\;{\cal M}(p_{1},...,p_{n};\mu ^{2})\,,  \label{eq:g5}     
\end{equation}     
where ${\cal M}(p_{1},...,p_{n};\mu ^{2})$ is given by (\ref{eq:ubsa}) and  
\begin{eqnarray}     
G^{\pm }&=&G_{0}^{\pm }\exp \left( -\frac{1}{2}W_{A}({\mbox{\boldmath     
$k$}_{\perp }^{2}},M^{2})-\frac{1}{2}w_{W}({\mbox{\boldmath     
$k$}_{\perp }^{2}%
},\mu ^{2})+\frac{1}{2}w_{W}({\mbox{\boldmath $k$}_{\perp }^{2}}%
,M^{2})\right) \,, \nonumber \\     
G^{Z}&=&G_{0}^{Z}+g\cos \theta _{w}T^{3}\left( \exp     
\left( -\frac{1}{2}W_{A}({%
\mbox{\boldmath $k$}_{\perp }^{2}},M^{2})\right) -1\right) \,,     
\nonumber \\
G^{\gamma }&=&G_{0}^{\gamma }+g\sin \theta _{w}T^{3}\left( \exp     
\left( -\frac{1%
}{2}W_{A}({\mbox{\boldmath $k$}_{\perp }^{2}},M^{2})\right) -1\right) \,.     
\label{eq:g6}     
\end{eqnarray}     
The important point here is the difference of the DL exponent for a     
nonradiative process and the process with photon emission, which leads to a     
violation of the Poisson distribution for photons in the DL approximation at     
high energies. It is a direct consequence of the fact that the photon has a     
non-Abelian component.     
     
\subsection{DL effects in semi-inclusive cross sections}     
     
Measurable cross sections have an inclusive nature (at least with respect to     
photons, since only cross sections with an infinite number of emitted soft     
photons are observable). Let us consider the DL corrections to such cross     
sections.     
     
It is clear that for the emission of real gauge bosons the same cut-off $\mu     
^{2} $ must be used as for virtual ones. Therefore, to calculate an     
experimentally measured cross section we have to take the cut-off $\mu ^{2}$     
less than the lower bound $\mu _{\exp}^{2}$ of ${\mbox{\boldmath $k$}_{\perp     
}^{2}}$ of those photons emitted in processes which are not included     
in the cross section.     
     
The calculation of the cross section is simple if the experimental     
conditions are such that only processes with emission of photons with ${%
\mbox{\boldmath $k$}_{\perp }^{2}}<M^ 
{2}$ are allowed. In this case the     
non-Abelian component of the photon is not essential, so that photon     
emissions obey a Poisson distribution. Therefore for the cross section     
with an arbitrary number of emitted photons with momenta lying inside regions     
$\Omega _{i}$ of the momentum space around the emitting particles with     
momenta $p_{i}$, we obtain:     
\[     
d\sigma (p_{1}, \ldots, p_{n}) = d\sigma _{\rm elastic}(p_{1}, \ldots ,p_{n}) \exp   
(w_{\exp}^{\gamma })\,,     
\]     
where $d\sigma _{\rm elastic}$ is the cross section of the non-radiative process     
and $w_{\exp}^{\gamma}$ is the probability of the emission of     
photons with ${\mbox{\boldmath $k$}_{\perp }^{2}}>\mu ^{2}$ inside the     
allowed region     
\begin{eqnarray}     
w_{\exp}^{\gamma }&=&\sum_{i=1}^{n}\frac{e_{i}^{2}}{(2\pi )^{3}}%
\int_{\Omega _{i}}\frac{d^{3}k}{2\omega }\frac{2E_{i}}{\omega (kp_{i})}%
\theta ({\mbox{\boldmath $k$}_{\perp }^{2}}-\mu ^{2}) \nonumber 
\\ &=&\sum_{i=1}^{n}\frac{%
e_{i}^{2}}{4\pi ^{3}}\int_{\Omega _{i}}\frac{d^{2}{\mbox{\boldmath $k$}%
_{\perp }}}{{\mbox{\boldmath $k$}_{\perp }^{2}}+m_i^2 \omega^2 / E_i^2}\frac{d\omega }{\omega }%
\theta ({\mbox{\boldmath $k$}_{\perp }^{2}}-\mu ^{2}).  \label{eq:h2}     
\end{eqnarray}     
Since the upper bound on ${\mbox{\boldmath $k$}_{\perp }^{2}}$ of the     
photons which are allowed to be radiated is less than $M^{2}$, we must use the     
cut-off $\mu ^{2}<M^{2}$ and, consequently, (\ref{eq:ubsa}) for the matrix element of  
the non-radiative process.  Therefore, we obtain     
\begin{eqnarray}     
d\sigma (p_{1},...,p_{n}) &=&d\sigma _{\rm Born}(p_{1},...,p_{n})  \nonumber \\     
&&\times \exp \left[ -\frac{g^{2}}{(4\pi )^{2}}\sum_{i=1}^{n}\left( \left\{     
T_{i}(T_{i}+1)+\tan ^{2}\theta _{w}\left( \frac{Y_{i}}{2}\right)     
^{2}\right\} \log ^{2}\frac{s}{M^{2}}\right) \right]  \nonumber \\     
&&\times \exp \left[ -\sum_{i=1}^{n}Q_{i}^{2}\left( w_{i}(s,\mu     
^{2})-w_{i}(s,M^{2})\right) +w_{\exp}^{\gamma }\right] \,,     
\label{eq:h3}     
\end{eqnarray}     
where $w_{i}(s,\mu ^{2})$ is given by (10) and (9). Evidently, the dependence on     
$\mu $ in $\sum Q_{i}^{2}w_{i}(s,\mu ^{2})$ and $w_{\exp}^{\gamma}$   
cancels in the exponential.     
     
If we include in the observed cross section the emission of gauge bosons     
with transverse momenta larger than $M$, then the problem becomes much more     
complicated because of the non-Poisson distribution of soft emission in     
non-Abelian gauge theories \cite{kf}. Let us consider here the simplest     
example of the cross section completely inclusive of photons emission in the     
two-loop approximation. Then the cross section can be written as     
\begin{equation}     
d\sigma (p_{1},...,p_{n})=d\sigma _{\rm Born}(p_{1},...,p_{n}) \left (1+\delta _{v}+%
\frac{\delta _{v}^{2}}{2}+\delta _{r}+\delta _{rv}+\frac{\delta     
_{r}^{2}}{2} \right )\,,  \label{eq:h4}     
\end{equation}     
where $\delta _{v}$ is the one-loop virtual correction, $\delta _{r}$ comes     
from one-photon emission taken in the Born approximation and $\delta _{rv}$ from the   
one-loop correction to one-photon emission. Due to     
exponentiation of the DL terms of the Sudakov-type in virtual corrections,     
the term $\delta _v^2/2$ in (\ref{eq:h4}) gives the two-loop     
virtual correction. The term $\delta _r^2/2$ gives the     
correction from two-photon emission in the Born approximation, since in     
this approximation the two photons are emitted independently.     
     
In the considered case of the cross section completely inclusive of photon     
emission the cut-off parameter $\mu ^{2}$ can be taken as large as $M^{2}$     
(but not greater, because the cross section does not include $W^{\pm }$ and     
$Z$ emission). Each of the corrections considered above depends on $\mu^{2}$, but their   
sum in (\ref{eq:h4}) does not; therefore, we can take the most     
suitable value of the cut-off to calculate the cross section. It is easy to     
see that the most convenient choice is $\mu ^{2}=M^{2}$. In this case from     
(26) we have:     
\begin{equation}     
\delta _{v}=-\frac{g^{2}}{(4\pi )^{2}}\sum_{i=1}^{n}\left( \left\{     
T_{i}(T_{i}+1)+\tan ^{2}\theta _{w}\left( \frac{Y_{i}}{2}\right)     
^{2}\right\} \log ^{2}\frac{s}{M^{2}}\right) \,.  \label{eq:h5}     
\end{equation}     
The correction due to one-photon emission taken in the Born approximation is   
$\delta _{r}=w_{\exp}^{\gamma }$ where $w_{\exp}^{\gamma }$ is     
given by (\ref{eq:h2}) with $\mu ^{2}=M^{2}$, $E_{i}\sim \sqrt{s}$ and the     
region $\Omega _{i}$ defined by inequality $\omega <E_{i}$. It gives     
\begin{equation}     
\delta _{r}=\sum_{i=1}^{n}Q_{i}^{2}w_{i}(s,M^{2})=\sum_{i=1}^{n}Q_{i}^{2}%
\frac{e^{2}}{(4\pi )^{2}}\log ^{2}\frac{s}{M^{2}}\,.  \label{eq:h6}     
\end{equation}     
The one-loop contribution to this correction $\delta _{rv}$ can be found     
with help of (26), (31), (32) and (33):     
\[     
\delta _{rv}=\sum_{i=1}^{n}\frac{e^{2}}{4\pi ^{2}}\int_{M^{2}}^{s}\frac{d{%
\mbox{\boldmath $k$}_{\perp }^{2}}}{{\mbox{\boldmath $k$}_{\perp }^{2}}}%
\int_{|{\mbox{\boldmath $k$}}_\perp|}^{\sqrt{s}}\frac{d\omega }{\omega }\left[     
Q_{i}^{2}\delta _{v}-Q_{i}T_{i}^{3}W_{A}({\mbox{\boldmath     
$k$}_{\perp }^{2}}%
,M^{2})\right]     
\]     
\begin{equation}     
=\sum_{i=1}^{n}\frac{e^{2}}{(4\pi )^{2}}\left[ \delta _{v}Q_{i}^{2}\log     
^{2}%
\frac{s}{M^{2}}-Q_{i}T_{i}^{3}\frac{g^{2}}{3(4\pi )^{2}}\log ^{4}\frac{s}{%
M^{2}}\right] \,.  \label{eq:h7}     
\end{equation}     
The important point here is that the virtual correction to the one-photon     
emission does not coincide with the corresponding correction to the     
non-radiative process (which means violation of the Poisson distribution) and     
depends on the momentum of the emitted photon. It is a consequence of the fact that the     
photon has a non-Abelian component.     
     
One can check that using an arbitrary cut-off $\mu ^{2}<M^{2}$ gives the     
same result, but the calculation is more complicated.     
     
\section{DL effects for amplitudes with Regge kinematics}     
     
For two particle scattering in either the forward $s\gg (-t)$ or the
backward $s\gg (-u)$ directions, corresponding to the Regge regime, there
are situations, where the amplitudes have DL contributions different from
those of the Sudakov type. These logarithms appear from Feynman diagrams in
which the soft particles with the minimal transverse momenta are pairs of
virtual fermions or bosons exchanged in the crossed $t$ or $u$ channels. For
QED and QCD the infrared evolution equations for such Regge processes are
known [4, 5, 7]. 
The new features presented by the Standard Model of weak
interactions are that the gauge group is semi-simple, $SU(2)\times U(1)$,
and that there is a large difference in the particle mass scales. The
general strategy of finding DL asymptotics in this case is to first solve
the evolution equation in the region $\mu \gg M_{1}$, where $M_{1}$ is the
largest particle mass in the theory; then to solve it in the region $%
M_{2}\ll \mu \ll M_{1}$, where $M_{2}$ is the next largest particle mass,
using the solution of the previous equation at $\mu =M_{1}$ as the initial
condition, and to proceed with these steps until the intermediate particles
are light quarks, electrons, neutrinos and photons. 

For simplicity we consider only the evolution equation in the region where
the cut-off parameter $\mu $ is much larger than all particle masses. In
this case it is natural to calculate the amplitudes in terms of the massless
particles of the unbroken theory, that is the leptons and quarks, the Higgs,
$B$ and $W^{a}$ bosons. By solving the generalized infrared evolution
equation in the DL approximation we sum up contributions of ladder-type
diagrams in the crossed channel, with all possible insertions of the gauge
bosons, leading to the Sudakov double logarithms. Because the transverse
momenta of the virtual particles are strongly ordered, the integral kernels
in the ladder diagrams are given by the splitting kernels of the DGLAP
evolution equ ation \cite{DGLAP} which describes all possible transitions
among fermions and bosons in the Standard Model. These splitting kernels can
be simplified since the Regge kinematic domain corresponds to a strong
ordering of the Bjorken variables $x_{i}$ along the ladder.

For definiteness we study the simple case of the backward lepton ($l$) -
antilepton ($\bar{l}$) scattering for $s\simeq -t\gg -u$. In the Born
approximation, the contribution of the Feynman diagrams with the Higgs,
photon, $B$ and $W^{a}$ boson exchanges depend on the helicities $\zeta
_{i}=\lambda _{i}/2$ (with $\lambda _{i}=\pm 1$) of the initial and final
fermions, which have momenta
\begin{equation}
p_{l},\;\;p_{\bar{l}},\;\;p_{l^{\prime }}\;\simeq
\;p_{\bar{l}},\;\;p_{\bar{%
l^{\prime }}}\;\simeq \;p_{l}.  \label{eq:xx}
\end{equation}
We consider the scattering of the leptons $e$ and $\nu _{e}$ belonging to
the first generation with the smallest masses. Because the coupling of the
Higgs particle to leptons is proportional to their mass, we can neglect the
Higgs contribution. The Dirac spinors describing the $l$ and $\overline{l}$
states with the definite helicities $\lambda $ are
\begin{equation}
u_{\lambda }(p)=\sqrt{\left| \mbox{\boldmath $p$}\right| }
\left( 
\begin{array}{l}
\varphi _{\lambda } \\ 
\lambda \,\varphi _{\lambda }
\end{array}
\right) \,,\quad \quad v_{\lambda }(-p)=\sqrt{\left| \mbox{\boldmath $p$}
\right| } \,\left( 
\begin{array}{l}
\varphi _{-\lambda } \\ 
-\lambda \varphi _{-\lambda }
\end{array}
\right) \,,\,\,\ 
\end{equation}
where 
\begin{equation}
\frac{\mbox{\boldmath $\sigma  p$}}{\left| \mbox{\boldmath $p$}\right| }
\,\varphi _{\lambda }=\lambda \,\varphi _{\lambda }.
\end{equation}
With the use of these expressions we can calculate the matrix elements of
the $\gamma $-matrix structures which appear in the Born approximation for
backward lepton-antilepton scattering, $l\overline{l} \rightarrow \overline{l%
}^\prime l^\prime$, 
\begin{eqnarray}
\frac{1}{s}\overline{v}_{\lambda _{\overline{l}}}(-p_{\overline{l}} )\gamma
_{\sigma }u_{\lambda _{l}}(p_{l})\,\overline{u}_{\lambda _{l^{\prime
}}}(p_{l^{\prime }})\gamma _{\sigma }v_{\lambda _{\overline{l^{\prime }}%
}}(-p_{\overline{l^{\prime }}})&\!\!=\!\!&a_{s}^{\lambda _{l}\,\lambda _{\overline{l}%
}\,\lambda _{l^{\prime }}\,\lambda _{\overline{l^{\prime }}}}=2\delta
_{\lambda _{l}\,,-\lambda _{\overline{l}}\,}\delta _{\,\lambda _{l^{\prime
}}\,,-\lambda _{\overline{l^{\prime }}}}\delta _{\lambda _{l}\,,-\lambda
_{l^{\prime }}}\,, \\
\,-\frac{1}{t}\overline{v}_{\lambda _{\overline{l}}}(-p_{\overline{l}%
})\gamma _{\sigma }v_{\lambda _{\overline{l^{\prime }}}}(-p_{\overline{%
l^{\prime }}})\,\overline{u}_{\lambda _{l^{\prime }}}(p_{l^{\prime }})\gamma
_{\sigma }u_{\lambda _{l}}(p_{l})&\!\!=\!\!&a_{t}^{\lambda _{l}\,\lambda _{\overline{l}%
}\,\lambda _{l^{\prime }}\,\lambda _{\overline{l^{\prime }}}}=2\delta
_{\lambda _{l}\,,\lambda _{\overline{l}}\,}\,\delta _{\,\lambda _{l^{\prime
}}\,,\lambda _{\overline{l^{\prime }}}}\,\delta _{\lambda _{l}\,,\lambda
_{l^{\prime }}}\,.
\end{eqnarray}
It is important, to note that both the helicity structures of $a_{s}$ and $%
a_{t}$ are proportional to $\delta _{\lambda _{l}\,,\lambda _{\overline{%
l^{\prime }}}\,} $, which means, that the helicities of the two fermions in
the crossed $u$ channel are opposite in sign. Therefore the Born diagram
with a virtual $W^{a}$ boson gives a negligible contribution to the backward
scattering. Because the $B$ boson interacts with hypercharge $Y$, its
contribution to the Born amplitude is also zero for the right-handed
neutrinos (if they were to exist).

Below we consider only the non-trivial case, when only
$e_{\lambda }^{\pm }$%
, $\nu _{-}$ and $\overline{\nu }_{+}$ can participate in the reaction.
Because the hypercharge $Y=-1$ for left-handed fermions and $Y=-2$ for the
right-handed electron, in this case the Born amplitude is given by
\begin{equation}
{\cal M}_{\mathrm{Born}}^{\lambda _{l}\,\lambda _{\overline{l}}\,\lambda
_{l^{\prime }}\,\lambda _{\overline{l^{\prime }}}}=\frac{g^{\prime 2}}{2}%
\,\left( a_{s}^{\lambda _{l}\,\lambda _{\overline{l}}\,\lambda _{l^{\prime
}}\,\lambda _{\overline{l^{\prime }}}}+\,a_{t}^{\lambda _{l}\,\lambda _{%
\overline{l}}\,\lambda _{l^{\prime }}\,\lambda _{\overline{l^{\prime }}%
}}\right) \,.
\end{equation}
In the DL approximation we also obtain a significant simplification of the
helicity structure of the scattering amplitude:
\begin{eqnarray}
{\cal M}^{\lambda _{l}\,\lambda _{\overline{l}}\,\lambda
_{l^{\prime }}\,\lambda _{\overline{l^{\prime }}}}&=&\frac{g^{\prime 2}}{2}%
\,\left( a_{s}^{\lambda _{l}\,\lambda _{\overline{l}}\,\lambda _{l^{\prime
}}\,\lambda _{\overline{l^{\prime }}}}f(s/\mu ^{2})+\,a_{t}^{\lambda
_{l}\,\lambda _{\overline{l}}\,\lambda _{l^{\prime }}\,\lambda _{\overline{%
l^{\prime }}}}f(-s/\mu ^{2})\right) \,, \\
\,f(s/\mu ^{2})&=&\,f^{+}(s/\mu ^{2})+f^{-}(s/\mu ^{2})\,,
\end{eqnarray}
where the functions $f^{\pm }(s/\mu ^{2})$ describe contributions with
positive and negative signature:
\begin{eqnarray}
f^{\pm }(-s/\mu ^{2}) &=&\pm f^{\pm }(s/\mu ^{2})\,,  \nonumber \\
f_{\mathrm{Born}}^{+}(s/\mu ^{2}) &=&1,\quad \quad f_{\mathrm{Born}%
}^{-}(s/\mu ^{2})=0.
\end{eqnarray}
The amplitudes $f^{\pm }$ are assumed to satisfy dispersion relations
of the form
\begin{equation}
f^{\pm }(s/\mu ^{2})=-\frac{1}{\pi }\int_{0}^{\infty }\left( \frac{s}{%
s-s^{\prime }}\pm \frac{s}{s+s^{\prime }}\right) \,\mbox{Im}\,f^{\pm
}(s^{\prime }/\mu ^{2})\frac{ds^{\prime }}{s^{\prime }}\,\,,
\end{equation}
where in the Born approximation $\mbox{Im}f_{
\rm Born}^{\pm }(x)$ is different
from zero only in the region where $x$ is of the order of unity. In
particular, from the known behavior of the Born
contribution $f_{\rm Born}(s/\mu ^{2})$ at large
$s$  and its degeneracy in the signature we
obtain
\[
-\frac{2}{\pi }\int_{0}^{\infty }\,\mbox{Im}f_{\rm Born}^{\pm}(s^{\prime}/\mu^2 )\,%
\frac{ds^{\prime }}{s^{\prime }}=1.
\]
The absence of degeneracy of the amplitude with respect to signature is a
result of the presence of the non-planar diagrams which arise from the
virtual soft $B$-bosons in the infrared evolution equations (cf. the QCD
case \cite{kl}):
\[
\frac{d\,f^{+}(s/\mu ^{2})}{d\log (\mu ^{2})}=-\frac{g^{\prime 2}}{8\pi
^{3}}%
\int_{0}^{\log \frac{s}{\mu ^{2}}}f^{+}(s_{1}/\mu ^{2})\,\,\left(
\int_{0}^{\log \frac{s}{s_{1}}}\mbox{Im}f^{+}(s_{2}/\,\mu ^{2})\,d\log
\frac{%
s_{2}}{\mu ^{2}}\right) \,d\log \frac{s_{1}}{\mu ^{2}}+
\]
\begin{equation}
\frac{\log ^{+}(-s/\mu ^{2})}{8\pi ^{2}}\,\left( \frac{3g^{2}+9g^{\prime
2}}{%
4}\right) \,\,f^{+}(s/\mu ^{2})+\,\frac{\log ^{-}(-s/\mu ^{2})}{8\pi ^{2}}%
\,\left( \frac{3g^{2}+g^{\prime 2}}{4}\right) \,f^{-}(s/\mu ^{2}),\,
\end{equation}
\[
\frac{d\,f^{-}(s/\mu ^{2})}{d\log (\mu ^{2})}=-\frac{g^{\prime 2}}{8\pi
^{3}}%
\int_{0}^{\log \frac{s}{\mu ^{2}}}f^{-}(s_{1}/\mu ^{2})\,\,\left(
\int_{0}^{\log \frac{s}{s_{1}}}\mbox{Im}f^{-}(s_{2}/\,\mu ^{2})\,d\log
\frac{%
s_{2}}{\mu ^{2}}\right) \,d\log \frac{s_{1}}{\mu ^{2}}+
\]
\begin{equation}
\frac{\log ^{+}(-s/\mu ^{2})}{8\pi ^{2}}\,\left( \frac{3g^{2}+9g^{\prime
2}}{%
4}\right) \,\,f^{-}(s/\mu ^{2})+\,\frac{\log ^{-}(-s/\mu ^{2})}{8\pi ^{2}}%
\,\left( \frac{3g^{2}+g^{\prime 2}}{4}\right) \,f^{+}(s/\mu ^{2})\,,
\end{equation}
where we introduced the notation

\[
\log ^{\pm }(-s/\mu ^{2})=\frac{\log (-s/\mu ^{2})\pm \log (s/\mu ^{2})}{2}%
\,.
\]

The linear terms in $f^{\pm }$, on the right-hand side of the evolution
equations, are the Sudakov contributions which arise from soft $W^{a}$ and
$%
B $ virtual bosons coupling to external lines with different momenta $%
p_{l}\simeq p_{\overline{l^{\prime }}}$ and $p_{l^{\prime }}\simeq p_{%
\overline{l}}$. The non-linear terms appear due to soft pairs of leptons
exchanged in the crossed $u$ channel.

Note that, in the above equations, the term related to the $s$-channel
Sudakov vertex with $W^{a}$ exchange is proportional to the Casimir operator
$T(T+1)=3/4$ for the weak isospin group $SU(2)$, and the total $B$ boson
contribution from the four Sudakov vertices having singularities in the $s$
and $t$ channels is proportional to $(-1-2)^{2}/4=9/4$. The corresponding
contribution of these vertices to the terms mixing the amplitudes with
different signatures is proportional to $3/4$ for $W^{a}$ exchange and $%
(-1+2)^{2}/4=1/4$ for $B$ exchange. Non-linear terms take into account the
contribution of the ladder diagrams for two leptons interacting through $B$
exchanges.

The above equations can be simplified if we use the Mellin representation
for the amplitudes:
\begin{equation}
f^{\pm }(\frac{s}{\mu^2})=\int_{L}\frac{dj}{4i}\,\left( \frac{s}{\mu ^{2}}\right)
^{j}\,\xi ^{\pm }(j)\,f_{j}^{\pm },
\end{equation}
where $\,\xi ^{\pm }(j)$ are the signature factors
\[
\xi ^{\pm }(j)=\frac{\exp (-i\pi \,j)\pm 1}{\sin (\pi j)}\,,
\]
and where the integration contour $L$ is situated along the imaginary axes
to the right of all singularities of the $t$-channel partial waves $%
f_{j}^{\pm }$. 
In the $j$-representation, taking into account the degeneracy of the partial
waves with different signatures in the Born approximation, the evolution
equations can be written in the 
form of Ricatti equations [7]:
\begin{eqnarray}
f_{j}^{+} &=&1-\frac{g^{\prime 2}}{16\pi ^{2}}\frac{1}{j^{2}}\left(
\,f_{j}^{+}\right) ^{2}+\left( \frac{3g^{2}+9g^{\prime 2}}{32\pi ^{2}}%
\right) \,\frac{d}{dj}(f_{j}^{+}/j)\,, \\
&&  \nonumber \\
f_{j}^{-} &=&1-\frac{g^{\prime 2}}{16\pi ^{2}}\frac{1}{j^{2}}\left(
\,f_{j}^{-}\right) ^{2}+\left( \frac{3g^{2}+9g^{\prime 2}}{32\pi ^{2}}%
\right) \,\frac{1}{j}\frac{df_{j}^{-}}{dj}\,-\left( \frac{3g^{2}+g^{\prime
2}%
}{32\pi ^{2}}\right) \,\frac{f_{j}^{+}}{j^2}\,,\,
\end{eqnarray}
if we take into account only the terms with effective DL variables
\begin{equation}
\frac{1}{\omega ^{2}}=\frac{g^{\prime 2}}{16\pi ^{2}}\,\frac{1}{j^{2}}%
\,,\quad \quad a=-\frac{2\,g^{\prime 2}}{3g^{2}+9g^{\prime 2}}\,
\end{equation}
and neglect other terms. Furthermore, using the following definition for the
$t$%
-channel partial waves
\begin{equation}
\varphi ^{\pm }(\omega )=f_{j}^{\pm }\,,
\end{equation}
the equations can be written as follows
\begin{eqnarray}
\varphi ^{+}(\omega )&=&1-\left( \,\frac{\varphi ^{+}(\omega )\,}{\omega }%
\right) ^{2}-\frac{1}{a}\,\frac{d}{d\omega }\left( \frac{\varphi ^{+}(\omega
)}{\omega }\right) \,\,, \\
\varphi ^{-}(\omega )&=&1-\left( \,\frac{\varphi ^{-}(\omega )}{\omega }%
\right) ^{2}-\frac{1}{a}\frac{1}{\omega }\,\frac{d\varphi ^{-}(\omega )}{%
d\omega }+\frac{1+4a}{a}\frac{\varphi
^{+}(\omega )%
}{\omega^2 } \,.
\end{eqnarray}
Both equations can be reduced to linear form after introducing new functions
$\psi (\omega )$ and $\chi (\omega )$ according to the definitions
\begin{equation}
\frac{\varphi ^{+}(\omega )\,}{\omega }=\frac{d\psi (\omega )/d\omega }{%
a\,\psi (\omega )}-\frac{\omega }{2}\,,\quad \quad \frac{\varphi ^{-}(\omega
)}{\omega }=\frac{d\chi (\omega )/d\omega }{a\,\chi (\omega )\,}-\frac{%
\omega }{2}-\frac{1}{2a\,\omega }\,.
\end{equation}
They have the form of Schr\"{o}dinger equations
\begin{eqnarray}
&& \!\!\!\!\!\! \left( -\frac{d^{2}}{d\omega ^{2}} + \frac{a^{2}}{4}\omega
^{2}+a^{2}+\frac{a}{%
2}\right) \psi (\omega )\!=\!0\,, \\
&& \!\!\!\!\!\! \left( -\frac{d^{2}}{d\omega ^{2}} + \frac{a^{2}}{4}\omega ^{2}-\frac{1}{%
4\omega ^{2}}+a+a^{2}+(1+4a)\,\left( \frac{1}{\omega }\left( \frac{d\psi
(\omega )/d\omega }{\psi (\omega )}\right) -\frac{a}{2}\right) \right) \chi
(\omega )\!=\!0\,.
\end{eqnarray}
The first is an equation for the harmonic oscillator. Therefore taking into
account the boundary condition
\begin{equation}
\lim_{\omega \rightarrow +\infty }\psi (\omega )\sim \exp
\left( \frac{a}{4}%
\omega ^{2}\right) \,\,\omega ^{a}\,,
\end{equation}
so as to match onto perturbation theory, we obtain the solution
\begin{equation}
\,\psi (\omega )=D_{a}(\sqrt{-a}\,\omega )\,,
\end{equation}
where $D_{p}(z)$ is the parabolic cylinder function:
\begin{equation}
\,D_{p}(z)=\frac{\exp (-z^{2}/4)}{\Gamma
(-p)}\int_{0}^{\infty }e^{-zx-\frac{%
x^{2}}{2}}x^{-p-1}dx\,.
\end{equation}
Therefore for the scattering amplitude with positive signature we have,
to DL accuracy,
\begin{equation}
\,f^{+}(s/\mu ^{2})=\int_{L}\frac{dj}{2\pi i}\,\left( \frac{s}{\mu ^{2}}%
\right) ^{j}\,\frac{1}{a}\frac{d}{dj}\log \left( \exp (-a\omega^2 /4)D_{a}(%
\sqrt{-a}\,\omega )\right) ,
\end{equation}
with $\omega =4\pi j/g^{\prime }$. The equation for negative signatures
can be easily solved for $a=-1/4$ (corresponding to complex values of the
Weinberg angle):
\begin{equation}
\chi (\omega )=\sqrt{\omega }\,\exp (-\omega ^{2}/16)\,\,\Psi
\left( \frac{1%
}{8},\,1,\,\frac{\omega ^{2}}{8}\right) \,,
\end{equation}
where $\Psi (\alpha ,\,\gamma ,\,z)$ is the confluent hypergeometric
function:
\begin{equation}
\,\Psi (\alpha ,\,\gamma ,\,z)=\frac{1}{\Gamma (\alpha )}\int_{0}^{\infty
}e^{-zx}x^{\alpha -1}(1+x)^{\gamma -\alpha -1}dx\,\,.
\end{equation}
Therefore for $a=-1/4$, the amplitude with negative signature, in the DL
approximation, is
\begin{equation}
\,f^{-}(s/\mu ^{2})=\frac{-i\pi }{2}\int_{L}\frac{dj}{2\pi
}\,\left( \frac{s%
}{\mu ^{2}}\right) ^{j}\,\,\frac{1}{a}\,j\,\frac{d}{dj}\,\,\log \,\Psi
\left( \frac{1}{8},\,1,\,\frac{\omega ^{2}}{8}\right) \,,
\end{equation}
with $\omega =4\pi j/g^{\prime }$. For general values of the parameter $a$
the solution can be obtained as a perturbation series in $\omega ^{-1}$.
This perturbative expansion is convergent for all values of the DL parameter
$g^2 \log ^{2}(s/\mu ^{2})$.

Because the function $D_{p}(z)$ does not have any singularities in the
complex plane, the high energy asymptotics of the scattering amplitude with
positive signature
\begin{equation}
f^{+}(s/\mu ^{2})\sim s^{j_{i}}\,,\quad \quad j_{i}=\frac{g^\prime}{4\pi}%
\frac{z_{i}}{\sqrt{-a}}
\end{equation}
is determined by the position of its zeroes, $D_{p}(z_{i})=0$, in the
left-half plane, Re$%
z_{i}<0$. The function $\Psi (\alpha ,\,1,\,z)$ has a singularity $\sim \ln
z
$ as $z\rightarrow 0$, and therefore the scattering amplitude with negative
signature behaves as
\begin{equation}
f^{-}(s/\mu ^{2})\sim \ln ^{-1}s
\end{equation}
at high energies. Such behavior of $f^{-}(s/\mu ^{2})$ is valid also for
other values of the parameter $a$. Note, that similar DL asymptotics were
obtained in QED for $e^{+}e^{-}$ backward scattering amplitudes. For
energies smaller than the masses of $W^{a}$ and $Z$ bosons, only diagrams
with virtual photons give rise to a DL contribution. The photons interact
with electric charge $Q=T_{3}+Y/2$, which is the same for left- and
right-handed electrons. It is interesting, that at very large energies, the
behavior of $e^{+}e^{-}$ backward scattering amplitudes is different from
the QED prediction even in the case of the Abelian Standard Model with
$g=0.$

The DL asymptotics of lepton-antilepton forward scattering amplitudes, which
is related to $t$-channel exchange of two fermions and an arbitrary number
of $W^{a}$ and $B$ bosons, can be obtained in a similar way. In this case
one should take into account the diagram with a virtual $W^{a}$ boson even
in the Born approximation. The reggeization of the fermions and $W^{a}$
boson can also be verified in the DL approximation with the use of the
infrared evolution equation. Moreover, one can construct amplitudes with
quasi-elastic unitarity \cite{qe}, which is important for the theory of the
BFKL Pomeron \cite{BFKL}. Some other applications of the DL asymptotics in
QCD are given in \cite{g1}, \cite{kmr}.

\section{Discussion}
  
We have calculated the double logarithmic (DL) corrections to the amplitudes
of a number of different high energy processes, in particular, of
electroweak processes in the Standard Model. These Sudakov-type corrections,
which are found to exponentiate, are crucial for the high precision studies
planned at the Next Linear $(e^{+}e^{-})$ Collider. Our approach is based on
the use of an evolution equation in the infrared cut-off parameter $\mu $,
which in turn is based on a generalization of a gauge invariant dispersive
method for photon Bremsstrahlung originally due to Gribov.
    
We have assumed that the energy of the process is much larger than all the
masses in the theory. However New Physics could appear in the TeV region and
it will modify our DL asymptotic forms. Because our approach is based on the
infrared evolution equation, we need simply to change the initial conditions
in this region to incorporate the new particles and/or new interactions of
the existing particles. The infrared evolution equation has the form of a
renormalization group equation in the two-dimensional impact parameter
subspace.
      
The usual BFKL equation can be also considered as a simplified version of
the renormalization group equation but in two-dimensional longitudinal
subspace. In this case, in the leading logarithmic approximation $g^{2}\log
s\sim 1$, instead of their dependence from the infrared cut-off $\mu$, we
have conformal invariance of the $t$-channel partial waves in impact
parameter space. However, in the next-to-leading approximation, the effect
of the running coupling leads to violation of conformal invariance
\cite{NLC}. After the breaking of conformal invariance the equation takes the form of
a quantized Callan-Symanzik equation \cite{conf}. Therefore, even in the
case of the BFKL equation, the renormalization group has its peculiarities.   
It is related to the fact, that the equation determines not only the   
anomalous dimensions of the operators of different twists, but also their   
relative contributions.   
   
The expressions for the DL asymptotics that we have obtained in the Standard   
Model can be used in perturbation theory to verify the first order, and to   
predict the higher order, expressions for scattering amplitudes, as has been   
done for the DL asymptotics in QCD \cite{g1}, \cite{BV}.  Finally, we reiterate that 
the accurate calculation of scattering amplitudes is important because the effects of   
New Physics can be rather small. \newline   

\section*{Added Note}
After this work was completed a paper by P. Ciafaloni and 
D. Comelli \cite{cc} appeared, in which 
the electroweak DL corrections are considered in the 
particular case of massless  $l\bar l$ production by a source which is a
singlet with respect to the $SU(2)\times U(1)$ gauge group. The results 
of this paper (in particular, nonexponentiating DL corrections) 
are not in agreement with our corresponding results.  
The difference appears due to the fact that
the method in \cite{cc}
consists of factorizing off the virtual gauge boson with the 
softest (i.e. with smallest frequency) momentum $k^{\mu}$ by 
computing external line insertions only, and in iterating this 
procedure by setting the virtual momentum $k=0$ in the left-over 
diagram.
We believe that
this approach, which is in disagreement with the method based
on the infrared evolution equation, is not valid.
  
\section*{Acknowledgements}  
   
We thank INTAS (95-0311) and the Royal Society for support. We also thank   
Ms. Sharon Fairless for help in preparing the manuscript.

\end{document}